\documentclass[aps,10pt,prb,twocolumn,letterpaper,superscriptaddress]{revtex4}

\usepackage{amsmath}
\usepackage{epsfig}
\usepackage{graphicx}
\usepackage{ragged2e}
\usepackage[justification=justified, format=plain]{caption}
\usepackage{latexsym}
\usepackage{amssymb}
\usepackage{amsfonts}
\usepackage{bm}
\usepackage{upgreek}

\newcommand{\be}{\begin{eqnarray}}
\newcommand{\ee}{\end{eqnarray}}
\newcommand{\bea}{\begin{eqnarray}}
\newcommand{\eea}{\end{eqnarray}}
\newcommand{\bpm}{\begin{pmatrix}}
\newcommand{\epm}{\end{pmatrix}}

\renewcommand{\v}[1]{{\boldsymbol{#1}}}

\begin{document}

\title{Superconducting fluctuations in overdoped Bi$_2$Sr$_2$CaCu$_2$O$_{8+\delta}$}

\author{Yu~He*}
\affiliation{Department of Applied Physics, Stanford University, Stanford, CA 94305, USA}
\affiliation{Department of Physics, University of California at Berkeley, Berkeley, CA 94720, USA}
\affiliation{Stanford Institute for Materials and Energy Sciences, SLAC National Accelerator Laboratory, 2575 Sand Hill Road, Menlo Park, CA 94025, USA}
\affiliation{Materials Sciences Division, Lawrence Berkeley National Laboratory, Berkeley, CA 94720, USA}
\thanks{These authors contributed equally to this work.}

\author{Su-Di~Chen*}
\affiliation{Department of Applied Physics, Stanford University, Stanford, CA 94305, USA}
\affiliation{Stanford Institute for Materials and Energy Sciences, SLAC National Accelerator Laboratory, 2575 Sand Hill Road, Menlo Park, CA 94025, USA}
\thanks{These authors contributed equally to this work.}

\author{Zi-Xiang~Li*}
\affiliation{Department of Physics, University of California at Berkeley, Berkeley, CA 94720, USA}
\thanks{These authors contributed equally to this work.}

\author{Dan Zhao}
\affiliation{Hefei National Laboratory for Physical Sciences at the Microscale, University of Science and Technology of China, Hefei, Anhui 230026, China}
\affiliation{CAS Key Laboratory of Strongly-coupled Quantum Matter Physics, Department of Physics, University of Science and Technology of China, Hefei, Anhui 230026, China}

\author{Dongjoon~Song}
\affiliation{National Institute of Advanced Industrial Science and Technology, Tsukuba 305-8565, Japan}

\author{Yoshiyuki~Yoshida}
\affiliation{National Institute of Advanced Industrial Science and Technology, Tsukuba 305-8565, Japan}

\author{Hiroshi~Eisaki}
\affiliation{National Institute of Advanced Industrial Science and Technology, Tsukuba 305-8565, Japan}

\author{Tao Wu}
\affiliation{Hefei National Laboratory for Physical Sciences at the Microscale, University of Science and Technology of China, Hefei, Anhui 230026, China}
\affiliation{CAS Key Laboratory of Strongly-coupled Quantum Matter Physics, Department of Physics, University of Science and Technology of China, Hefei, Anhui 230026, China}

\author{Xian-Hui Chen}
\affiliation{Hefei National Laboratory for Physical Sciences at the Microscale, University of Science and Technology of China, Hefei, Anhui 230026, China}
\affiliation{CAS Key Laboratory of Strongly-coupled Quantum Matter Physics, Department of Physics, University of Science and Technology of China, Hefei, Anhui 230026, China}

\author{Dong-Hui~Lu}
\affiliation{Stanford Synchrotron Radiation Lightsource, SLAC National Accelerator Laboratory, Menlo Park, CA 94025, USA}

\author{Christoph Meingast}
\affiliation{Institute for Quantum Materials and Technologies, Karlsruhe Institute of Technology, 76021 Karlsruhe, Germany}


\author{Thomas P. Devereaux}
\affiliation{Department of Materials Science and Engineering, Stanford University, Stanford, CA 94305, USA}
\affiliation{Stanford Institute for Materials and Energy Sciences, SLAC National Accelerator Laboratory, 2575 Sand Hill Road, Menlo Park, CA 94025, USA}

\author{Robert J. Birgeneau}
\affiliation{Department of Physics, University of California at Berkeley, Berkeley, CA 94720, USA}
\affiliation{Materials Sciences Division, Lawrence Berkeley National Laboratory, Berkeley, CA 94720, USA}

\author{Makoto~Hashimoto$^\dagger$}
\affiliation{Stanford Synchrotron Radiation Lightsource, SLAC National Accelerator Laboratory, Menlo Park, CA 94025, USA}

\author{Dung-Hai Lee$^\dagger$}
\affiliation{Department of Physics, University of California at Berkeley, Berkeley, CA 94720, USA}
\affiliation{Materials Sciences Division, Lawrence Berkeley National Laboratory, Berkeley, CA 94720, USA}

\author{Zhi-Xun~Shen$^\dagger$}
\affiliation{Department of Applied Physics, Stanford University, Stanford, CA 94305, USA}
\affiliation{Stanford Institute for Materials and Energy Sciences, SLAC National Accelerator Laboratory, 2575 Sand Hill Road, Menlo Park, CA 94025, USA}


\begin{abstract}
Fluctuating superconductivity - vestigial Cooper pairing in the resistive state of a material - is usually associated with low dimensionality, strong disorder or low carrier density. Here, we report single particle spectroscopic, thermodynamic and magnetic evidence for persistent superconducting fluctuations in heavily hole-doped cuprate superconductor Bi$_2$Sr$_2$CaCu$_2$O$_{8+\delta}$ ($T_c$ = 66~K) despite the high carrier density. With a sign-problem free quantum Monte Carlo calculation, we show how a partially flat band at ($\pi$,0) can help enhance superconducting phase fluctuations. Finally, we discuss the implications of an anisotropic band structure on the phase-coherence-limited superconductivity in overdoped cuprates and other superconductors.
\end{abstract}

\maketitle

\section{Introduction}

\begin{figure}
	\captionsetup{width=1\columnwidth,justification=RaggedRight}
	\includegraphics[width=1\columnwidth]{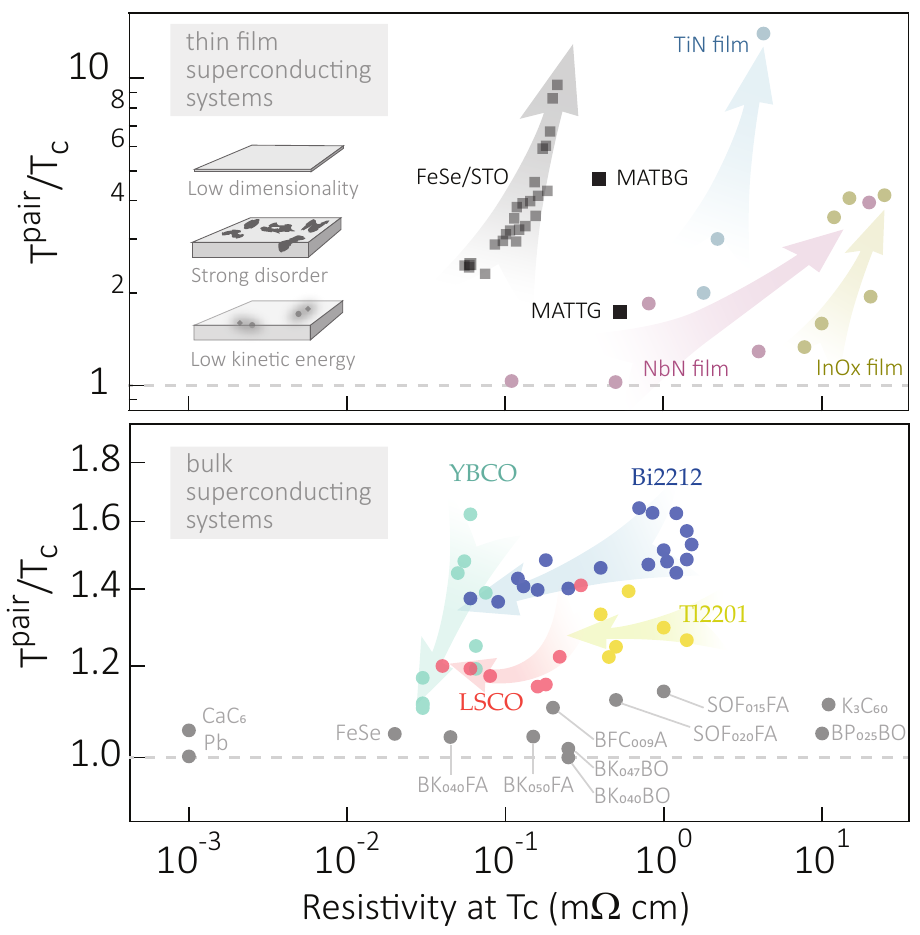}
	\caption{Thermal superconducting fluctuation effect in conventional and unconventional superconductors. Superconducting gap opening temperature $T_\textrm{pair}$ divided by zero resistance temperature $T_c$ tallied according to the resistivity right above $T_c$ for (top) thin film superconductors and (bottom) bulk superconductors. Arrows denote increasing disorder in thin film systems, and increasing hole doping in the cuprates. Except for TBG and TTG systems, $T_\textrm{pair}$ is exclusively determined from heat capacity or single-particle spectra~\cite{supp2021flucSC}. MATBG (MATTG) - magic angle twisted bilayer (trilayer) graphene. FeSe/STO - monolayer FeSe on Nb:SrTiO$_3$. BKFA - Ba$_{1-x}$K$_x$Fe$_2$As$_2$. BFCA - Ba(Fe$_{1-x}$Co$_x$)$_2$As$_2$. BKBO - Ba$_{1-x}$K$_x$BiO$_3$. BPBO - BaPb$_{1-x}$Bi$_x$O$_3$. SOFFA - SmO$_{1-x}$F$_x$FeAs. YBCO - YBa$_2$Cu$_3$O$_{7-\delta}$. Tl2201 - Tl$_2$Ba$_2$CuO$_{6+\delta}$. LSCO - (La,Sr)$_2$CuO$_4$. Bi2212 - Bi$_2$Sr$_2$CaCu$_2$O$_{8+\delta}$~\cite{supp2021flucSC}.} 
	\label{fig:figure0}
\end{figure}

In conventional superconductors, the superconducting (SC) transition temperature $T_c$ is controlled by the formation of Cooper pairs via exchange of low-energy pairing bosons of energy $\Omega_p$ in the BCS paradigm~\cite{bardeen1957theory}. Here, the normal state has many high kinetic energy carriers ($\epsilon_\textrm{F} \gg \Omega_p$), and pairing happens only between quasiparticle eigenstates with well-defined energy and momentum on a thin shell around the Fermi surface. Meanwhile, with the fluctuations in the pairing field neglected, superconducting phase coherence and Cooper pairing are simultaneously established via a mean-field second order phase transition at $T_c$. Thermal superconducting fluctuation is only notable near $T_c$ within a small temperature range $\delta T$, which is dictated by the Ginzburg criteria~\cite{ginzburg1961some}. In a three dimensional clean metal~\footnote{\textit{Clean} here means the normal state mean free path $l_\textrm{MFP}$ is larger than the superconducting coherence length $\xi_\textrm{BCS}$}, $\delta T/T_c \ll 1$ is expected theoretically and observed experimentally ~\cite{shiffman1963specific,hirshfeld1962superconducting,rorer1963specific}.


Investigations of the superconductor-to-insulator transition in superconducting thin films reveal prevailing fluctuations that are largely controlled by disorder and dimensionality~\cite{sacepe2010pseudogap,mondal2011phase,hao2021electric,cao2020strange,faeth2020incoherent}. There, higher level of disorder (measured by residual resistivity) can rapidly increase the separation between the superconducting pairing temperature $T_\textrm{pair}$ and the zero resistance temperature $T_c$. The thermodynamic superconducting transition happens at $T_c$, where superconducting correlation length diverges, while $T_\textrm{pair}$ becomes a crossover temperature scale that signifies the occurrence of non-zero superconducting pairing field modulus~\cite{larkin2005theory}. Between $T_c$ and $T_\textrm{pair}$, the phase of the superconducting order parameter remains disordered. Due to the crossover nature, experimental measures of $T_\textrm{pair}$ may depend on the nature of the probe. In thermodynamic probes such as heat capacity, $T_\textrm{pair}$ may be represented as the ``mean field'' pairing temperature $T_\textrm{MF}$ via equal entropy reconstruction~\cite{tallon2011fluctuations}; whereas in single particle probes, it is often signified by the spectral gap opening temperature $T_\textrm{gap}$~\cite{chen2019incoherent}. Figure.~\ref{fig:figure0}(a) aggregates the severity of superconducting fluctuations ($T_\textrm{pair}/T_c$) for most commonly studied 2D superconductors as a function of materials' resistivity at $T_c$, which is considered a measure of disorder within the same material family. It should be noted that for strictly mono-to-few atomic layer systems such as FeSe/SrTiO$_3$ or twisted graphene, $T_c$ is always suppressed from $T_\textrm{pair}$ due to prevalent vortex excitations even in the absence of disorder~\cite{kosterlitz1973ordering}. Clearly in these systems, fluctuating superconductivity can survive up to a $T_\textrm{pair}$ several times higher than $T_c$. In contrast, bulk superconducting systems exhibit fluctuations to a much lesser degree (Fig.~\ref{fig:figure0}(b)), which is consistent with the Ginzburg criteria in mean field superconducting transitions. Despite a range of over 4 decades of resistivity at $T_c$ in different compounds, high-$T_c$ cuprates stand out with the most prominent superconducting fluctuations, by a large margin. Within the four listed cuprate families, superconducting fluctuation in Bi$_2$Sr$_2$CaCu$_2$O$_{8+\delta}$ (Bi-2212) is the most prominent (Fig.~\ref{fig:figure0}(b)).

Strong superconducting fluctuation has been interpreted as a consequence of quasi-2-dimensionality and correlation induced low carrier density in the underdoped (UD) to optimally doped (OP) cuprates~\cite{corson1999vanishing,bollinger2011superconductor,uemura1989universal,emery1995importance}. In these systems, carrier kinetic energy is low due to strong on-site Coulomb repulsion, and an in/out-of-plane resistive anisotropy as high as $\rho_\textrm{c}/\rho_\textrm{ab}\sim10^5-10^6$ has been observed~\cite{ono2003evolution,chen1998anisotropic}. Both aspects jeopardize the mean field premise of the BCS theory, and provide basis for superconducting fluctuations. In this regime, it is the condensation of Cooper pairs - establishment of phase coherence - that determines $T_c$. Moreover, the normal state of UD and OP cuprates exhibits pseudogap and strange metal behaviors due to strong electronic correlation. It no longer hosts quasiparticles with well-defined energy-momentum relation stretching the entirety of the Fermi surface, undermining the Cooper instability necessary for BCS pairing~\cite{maier2016pairing}. More hole doping is shown to gradually restore quasiparticles to the normal state from the nodal direction towards the antinodal direction~\cite{sobota2020electronic}. After a critical doping $p_c$ is surpassed, the normal state regains spectral coherence~\cite{chen2019incoherent}, mode-coupling signature becomes rapidly weakened~\cite{he2018rapid}, and the superconducting gap-to-$T_c$ ratio starts to evolve towards the weak-coupling \textit{d}-wave BCS limit~\cite{sobota2020electronic}.

Indeed, transport~\cite{ando2004electronic,proust2002heat}, thermodynamic~\cite{tallon2011fluctuations} and single particle probes~\cite{yusof2002quasiparticle, chuang2004bilayer, chen2019incoherent, fujita2014simultaneous} all seem to point towards a more 3-dimensional, carrier-rich metallic normal state in overdoped (OD) cuprates. This leads to a widely held belief that the BCS paradigm has a chance to succeed here. Therefore, it is quite a surprise when recent measurements show that it is also phase coherence that determines $T_c$ in overdoped (La,Sr)$_2$CuO$_4$ (LSCO)~\cite{rourke2011phase,bovzovic2016dependence,mahmood2019locating}. Scanning tunneling spectroscopy (STS) indicates substantial nanoscale electronic inhomogeneity in overdoped Bi-2212, signifying the important role of disorder~\cite{gomes2007visualizing,parker2010nanoscale}. Photoemission evidence of fluctuating superconductivity have been observed in optimally doped~\cite{kondo2011disentangling} and overdoped cuprates~\cite{kondo2015point}, albeit mostly near the more coherent nodal direction. It remains unclear whether such a residual gap above $T_c$ is truly of superconducting origin, or is a vestigial presence of the incoherent pseudogap and/or more exotic gapped states~\cite{vishik2012phase,hashimoto2015direct,drozdov2018phase,chen2019incoherent,dai2020modeling}. The ambiguity on the nature of the antinodal spectral gap above $T_c$ in overdoped cuprates inadvertently created incompatible boundaries trisecting the pseudogap, superconductivity, and normal metal phases~\cite{kordyuk2015pseudogap}. Last but not least, concerted electrical, magnetic, thermodynamic and single-particle measurements on the same overdoped samples have been lacking, lending much uncertainty to conclusions pieced together from different cuprate families.

Motivated by these puzzles, we carry out a multi-technique investigation on one cuprate system Bi-2212, aiming to answer the following questions pertinent to overdoped cuprates and beyond:
\begin{enumerate}
    \item\vspace{-1 mm} Is there any residual ``pseudogap'' that is not of superconducting origin in overdoped cuprates;
    \item\vspace{-2 mm} How do the transport, thermodynamic, magnetic and single-particle properties of the normal and superconducting states come together on the same model system Bi-2212;
    \item\vspace{-2 mm} What may suppress the superfluid density and phase stiffness despite the high carrier density in the overdoped regime;
    \item\vspace{-2 mm} Will superconducting phase coherence imprint on the single particle spectral function in this case.
\end{enumerate}

\section{Material synthesis and characterization}

\begin{figure}
	\captionsetup{width=1\columnwidth,justification=RaggedRight}
	\includegraphics[width=1\columnwidth]{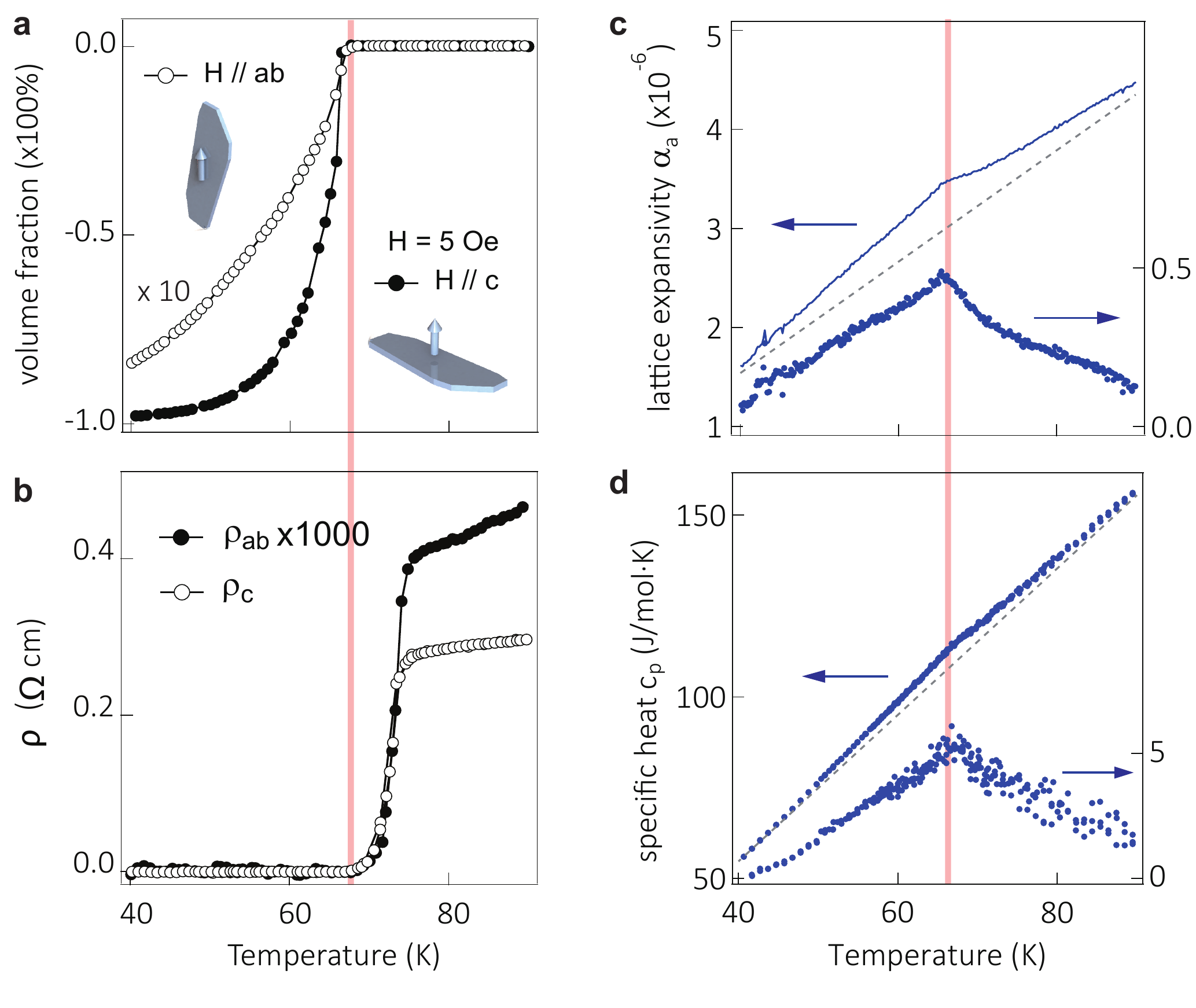}
	\caption{Bulk magnetic and thermodynamic properties of a heavily overdoped Bi-2212 with $T_c$ = 66~K. (a) DC diamagnetic response at 5 Oe, for fields both perpendicular (solid) and parallel (open) to the CuO$_2$-plane. (b) In- and out-of-plane resistivity near $T_c$. (c) In-plane linear lattice expansion coefficient $\alpha$. (d) Molar specific heat $c_p$. In both (c) and (d), a linear background (grey dash) is removed to highlight the superconducting transition in the bottom curve plotted to the right axes on an enlarged scale.}
	\label{fig:figure1}
\end{figure}

Overdoped bilayer cuprate (Pb,Bi)$_2$Sr$_2$CaCu$_2$O$_{8+\delta}$ (Pb:Bi-2212) is chosen for its excellent cleavability and large superconducting energy scale. Pressurized oxygen annealing from 2-400 bar is used to reach 66~K $> T_c >$ 51~K, which corresponds to nominal hole doping of $0.22 < p < 0.24$~\cite{presland1991general}. Figure~\ref{fig:figure1} shows the magnetic, electrical transport and thermodynamic signatures of a superconducting transition at $T_c =$ 66~K ($p$ = 0.22). A sharp superconducting transition defined by the Meissner diamagnetism (Fig.~\ref{fig:figure1}(a)) coincides to within 1~K with the zero-resistivity temperature (Fig.~\ref{fig:figure1}(b)) and the singularity in both the in-plane lattice expansivity (Fig.~\ref{fig:figure1}(c)) and heat capacity (Fig.~\ref{fig:figure1}(d)). A linear background is subtracted from the lattice expansivity and heat capacity data to highlight the transition~\footnote{A more involved background subtraction method was carried out by Ref.~\cite{tallon2011fluctuations,loram2004absence}, where a nonlinear phonon background from a Zn-doped non-superconducting sample was used. Despite the different background subtraction schemes, the extracted mean-field pairing temperature T$_\textrm{MF}$ is consistently 30-40\% higher than the resistive transition $T_c$.}. Compared to the optimally doped system~\cite{gu1998anisotropy,chen1998anisotropic}, here the out-of-plane and in-plane magnetization anisotropy $M_\textrm{c}/M_\textrm{ab}$ is reduced from 1400 to 12, while the corresponding resistivity anisotropy $\rho_\textrm{c}/\rho_\textrm{ab}$ is reduced from 10$^5$ to 600. This suggests a rapid restoration towards three dimensionality with 22\% hole doping in both the superconducting and normal states. The thermodynamic singularities are consistent with a sharp second order superconducting phase transition across $T_c$. But the extended critical region up to 20~K above $T_c$ signal substantial thermal fluctuations that cannot be rationalized within a simple BCS mean-field picture.

\section{Spectroscopic results}

\begin{figure*} 
	\captionsetup{width=2\columnwidth,justification=RaggedRight}
	\includegraphics[width=2\columnwidth]{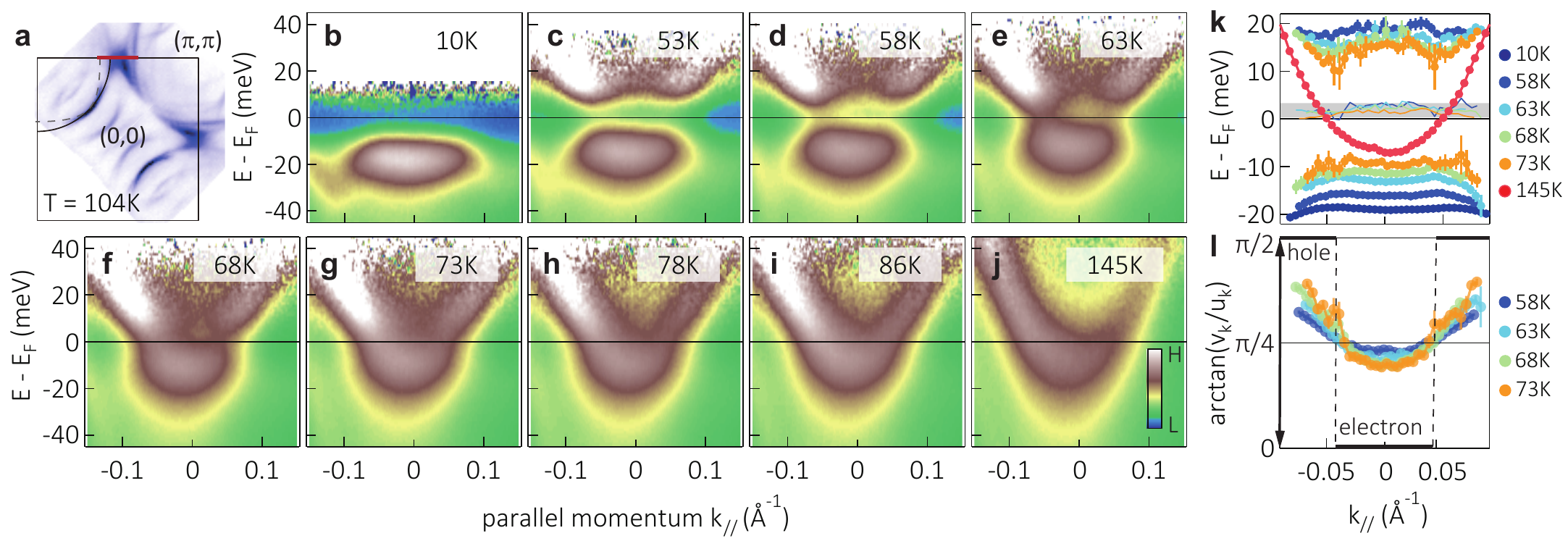}
	\caption{Temperature dependent high resolution antinodal energy-momentum spectra in heavily overdoped Bi-2212 with $T_c$ = 66~K. (a) Fermi surface map integrated within 10~meV of the chemical potential of a Pb-free sample in the normal state. Red bar denotes the parallel momentum of the spectra shown in subsequent panels. (b)-(j) Fermi-function-divided antinodal spectra along the BZ boundary in a Pb-doped sample. (k) Fitted energy-momentum dispersions of both the electron and hole Bogoliubov quasiparticles at various temperatures. Red circles indicate the normal state dispersion at 145~K. Thin lines shaded in grey are the averaged energies of the electron and hole Bogoliubov quasiparticle branches. (l) Calculated particle-hole mixing ratio, expressed in its arc tangent value, from the fitted intensity ratio of the electron and hole Bogoliubov quasiparticle branches near $T_c$. $\pi/4$ represents equal particle-hole mixture. Black line shows the BCS expectation when $\mathbf{v}_\text{F}\cdot\mathbf{k}_\text{F} \gg \Delta$.}
	\label{fig:figure2}
\end{figure*}

To obtain spectroscopic insights, we investigate the same system near the SC gap opening temperature ($T_\text{gap}$) and the zero-resistivity temperature ($T_c$) using high resolution angle-resolved photoemission spectroscopy (ARPES) and $^{63}$Cu nuclear magnetic resonance (NMR)~\cite{supp2021flucSC}. Systematic high resolution ARPES measurements in this temperature and doping regime is made possible by the recent advance in localized heating method~\cite{chen2019incoherent,kondo2015point}, which is instrumental to maintain the surface doping level in such super-oxygenated samples (see also Fig.~S\ref{fig:figureS2} for temperature cycles).

First, well-defined single particle spectral peaks are found over the untruncated normal state Fermi surface well above $T_c$ (Fig.~\ref{fig:figure2}(a)(j), Fig.~S\ref{fig:figureS1}), indicating a more metallic normal state compared to OP samples~\cite{chen2019incoherent}. Experimentally fitted Fermi surface volumes yield an averaged doping of $p$ = 0.26 between the bonding and antibonding sheets ($p_\text{AB}$ = 0.34, $p_\text{BB}$ = 0.18), slightly larger than the doping deduced from the empirical parabolic $T_c$-$p$ relation~\cite{presland1991general,drozdov2018phase}\footnote{It should be noted that the Luttinger volume determined this way may also deviate from the actual filling in the presence of lingering correlation effects~\cite{gros2006determining}.}. Subsequent discussions will mainly focus on the antibonding band, whose van Hove singularity is closest to the Fermi level.

To determine $T_\text{gap}$, a high statistics energy-momentum cut along the Brillouin zone (BZ) boundary (red line in Fig.~\ref{fig:figure2}(a)) is measured between 10~K and 150~K (Fig.~\ref{fig:figure2}(b)-(j)) with an energy resolution of 8~meV. The Fermi function is divided from the spectra to reveal the unoccupied states (for detailed procedures see Fig.~S\ref{fig:figureS2}~S\ref{fig:figureS14}; for raw EDCs see Fig.~S\ref{fig:figureS3}). Enabled by excellent statistics and resolution, both the electron and hole branches of the Bogoliubov quasiparticle dispersions can be traced out around $T_c$ (Fig.~\ref{fig:figure2}(k)). The superconducting gap decreases from 10~K up to 63~K, yet a sizeable spectral gap remains at $T_c$ (cyan markers). In fact, above $T_c$ it continues to close and fill up until it becomes indiscernible around 86~K. During this process, the particle-hole symmetry - evidenced by the dispersion (Fig.~\ref{fig:figure2}(k)) and intensity (Fig.~\ref{fig:figure2}(l)) of electron and hole branches of the Bogoliubov quasiparticles - remains preserved down to 1.5~meV uncertainty (grey band in Fig.~\ref{fig:figure2}(k)), which is quantitatively accountable by the finite resolution effect (Fig.~S\ref{fig:figureS14}). This particle-hole symmetric normal state gap suggests the presence of Cooper pairing beyond the nodal region~\cite{kondo2015point}, and is fundamentally different from the incoherent, particle-hole asymmetric pseudogap in the UD and OP cuprates (Fig.~S\ref{fig:figureS11})~\cite{hashimoto2010particle,chen2019incoherent}. A uniformly closing $d$-wave gap over the entire momentum space upon warming to T$_\textrm{gap}$ also disfavor non-zero momentum pairing in this temperature range (Fig.~S\ref{fig:figureS1}).

To quantify the particle-hole mixing near the antinodal region, Fig.~\ref{fig:figure2}(l) shows the Bogoliubov angle~\cite{fujita2008bogoliubov} $\theta_\mathbf{k}=\tan^{-1}|v_\mathbf{k}/u_\mathbf{k}|$ for temperatures near $T_c$. Here, $u_\mathbf{k}^2 (v_{\mathbf{k}}^2) $ are the quasiparticle spectral intensities below and above the Fermi energy. Accordingly, $\theta_\mathbf{k} =$ 0, $\pi/2$ and $\pi/4$ indicate pure electron (particle), pure hole, and equal particle-hole mixture respectively. For conventional superconductors $\theta_\mathbf{k}$ only deviates from 0 or $\pi/2$ in the vicinity of Fermi momenta $\mathbf{k}_F$, since $\Delta/E_F \ll 1$. However in overdoped Bi-2212 $\theta_\mathbf{k}$ is centered around $\pi/4$ for the entire Bogoliubov quasiparticle band around the antinode. We attribute this to the flatness of the normal state antinodal dispersion in comparison to the size of the superconducting gap.~\footnote{Without a superconducting gap, the van Hove point at ($\pi$,0) is only 7~meV below the Fermi energy. All of the electronic states here will be absorbed into the 16~meV superconducting gap well below $T_c$.}

\begin{figure}
	\captionsetup{width=1\columnwidth,justification=RaggedRight}
	\includegraphics[width=1\columnwidth]{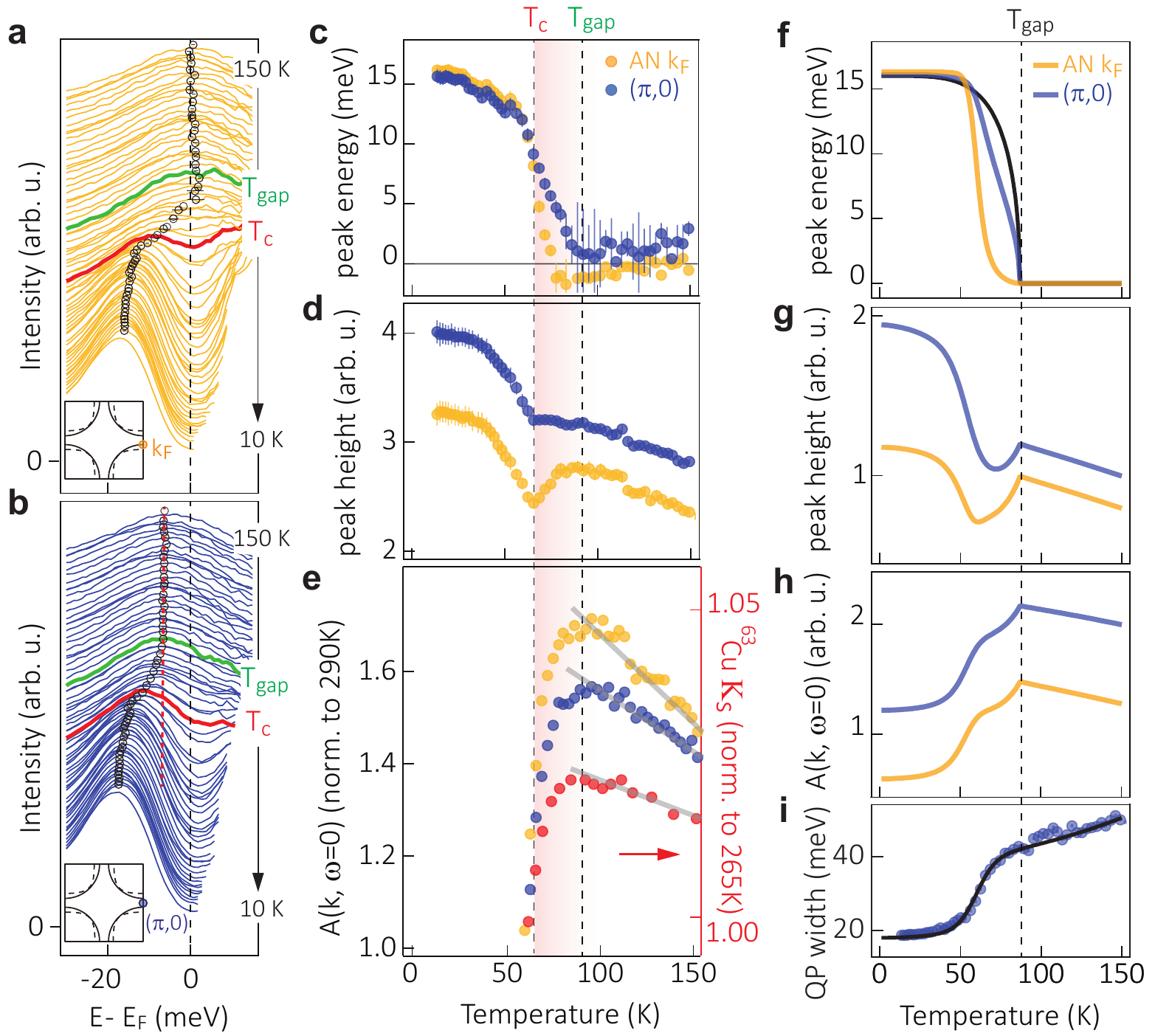}
	\caption{Temperature dependence of the antinodal EDCs. Fermi-function divided EDCs at (a) antinodal $\mathbf{k}_\text{F}$ and (b) ($\pi$,0). Insets depicts the two momenta in the Brillouin zone. Spectra are normalized to 1 over [0,100]~meV binding energy. Green and red lines denote $T_\textrm{gap}\sim$ 90~K and $T_c\sim$ 66~K respectively. Black circles mark the apparent spectral peak position, and the red-dashed line marks the ungapped van Hove point position. Temperature dependent (c) spectral peak binding energy, (d) spectral peak height, and (e) spectral intensity at $E_F$ from the antinodal $\mathbf{k}_\text{F}$ (yellow) and ($\pi$,0) (blue). Right axis in (e) corresponds to the frequency shift from the $^{63}$Cu nuclear spin resonance (red). Grey lines are guides to exemplify normal state spectral weight evolution. (f) Spectral peak binding energy, (g) spectral peak height, and (h) spectral intensity at $E_F$ extracted from the simulated spectra at the antinodal $\mathbf{k}_\text{F}$ (yellow) and ($\pi$,0) (blue), using the same analysis as (c)-(e). Black line in (f) is the BCS gap used for simulation, set to open at 87~K. The vertical shades in (c)-(e) indicate the superconducting fluctuation region. (i) Spectral peak linewidth fitted from ($\pi$,0) (blue), and the linewidth used for simulation (black). Curves in (d)(g)(h) are offset for clarity.}
	\label{fig:figure3}
\end{figure}

We then show that both $T_c$ and $T_\textrm{gap}$ can be uniquely determined from the same set of single-particle spectra. The electronic structure undergoes three distinct stages of evolution as a function of temperature. Figure ~\ref{fig:figure3}(a)(b) track the temperature dependence of the energy distribution curve (EDC) at the antinodal $\mathbf{k}_\text{F}$ and ($\pi$,0) on the antibonding band. \textbf{(1)} Cooling towards $\sim$90~K (green lines), the spectral peak gradually sharpens in width and grows in height (for the full temperature range up to 290~K, see Fig.~S\ref{fig:figureS10}(a)). \textbf{(2)} An energy gap gradually opens below $T_\textrm{gap}$. It starts as the zero-energy spectral intensity saturates, which then further splits into two largely overlapping peaks at $T_c$ (red lines). \textbf{(3)} Further cooling from $T_c$ instigates a rapid sharpening of the Bogoliubov quasiparticle peak, as well as a concomitant rapid depletion of the zero-energy spectral weight until a near-complete energy gap forms. Figure~\ref{fig:figure3}(c)(d)(i) quantitatively describe in the temperature evolution of the superconducting gap size, spectral peak height, and linewidth (minimally fitted by a Gaussian between -20~meV and 1~meV) at the antinodal $\mathbf{k}_F$ and ($\pi$,0). In particular, taking advantage of the shallow van Hove point, the superconducting gap at ($\pi$,0) may be extracted with excellent numerical stability via subtraction of quadrature $\Delta_\mathbf{k} = \sqrt{E_\mathbf{k}^2-\epsilon_\mathbf{k}^2}$ (Fig.~\ref{fig:figure3}(c), blue circles; see also Fig.~S\ref{fig:figureS10}). $E_\mathbf{k}$ and $\epsilon_\mathbf{k}$ are superconducting state and normal state quasiparticle dispersions.

While the momentum resolved single particle spectra most clearly show both $T_\text{gap}$ and $T_c$, the onset of zero-energy spectral weight depletion also signifies $T_\text{gap}$ (Fig.~\ref{fig:figure3}(e)). This is true both near the antinode shown by ARPES (blue and yellow circles), and as a momentum average shown by the Knight shift of $^{63}$Cu nuclear magnetic resonance (red circles). The fact that bulk probes, such as NMR, yield results consistent with the surface sensitive ARPES results further supports that the observed fluctuations are a bulk phenomenon~\footnote{Consistent ARPES and NMR signatures are also observed in 23\% and 24\% hole-doped Bi-2212 (Fig.~S\ref{fig:figureS130},~S\ref{fig:figureS8},~S\ref{fig:figureS7}).}.The spectral features shown in Fig.~\ref{fig:figure3}(a)-(e) can be excellently reproduced in a spectral simulation with only two inputs (Fig.~\ref{fig:figure3}(f)-(h)): a BCS-type $\Delta$(T) with a 16~meV zero-temperature BCS gap opening at $T_\text{gap}\sim87$~K (Fig.~\ref{fig:figure3}(f) black line), and a temperature-dependent linewidth (Fig.~\ref{fig:figure3}(i) black line) approximated from experimental measurements (Fig.~\ref{fig:figure3}(i) blue circles, see supplement for details). It is worth noting that spectral features around $T_c$ are caused by the sudden reduction of spectral linewidth instead of the gap opening. Clearly, there is a mismatch between the bulk gap-opening temperature $T_\text{gap}\sim$ 87~K and the bulk superconducting transition temperature $T_c=$ 66~K. Intriguingly, replacing $T_c$ with $T_\textrm{gap}$ in the superconducting gap-to-$T_c$ ratio $\frac{2\Delta(0)}{k_\text{B}T}$, this quantity finally reduce to the weakly-coupled \textit{d}-wave BCS value of $\sim$~4.3~\cite{he2018rapid}.

\section{Sign-problem free quantum Monte Carlo simulations}

So far, upon cooling, our data point to a rather conventional gap opening process intervened by strong phase fluctuations until coherence is achieved. This is consistent with the surprising observation of low zero temperature superfluid density in overdoped LSCO~\cite{bovzovic2016dependence}. However, the microscopic mechanism underlying the low superfluid density is unclear. Electronic inhomogeneity and disorder is a natural candidate to disrupt the phase coherence in this regime. But prevalent phase fluctuations in virtually all cuprate families, which have drastically different disorder levels, seem to suggest additional mechanisms~\cite{wang2007weak,bovzovic2019really}. Mean field models of dirty $d$-wave superconductors account for the reduction of zero-temperature superfluid density~\cite{lee2018optical,lee2020low}. However, the dirty limit is hard to reach with the extremely short superconducting coherence length~\cite{loram2004absence,mahmood2019locating}, and the model do not produce notable thermal fluctuations. Here we explore another factor which also contributes to the superfluid density reduction, namely, the flat dispersion near the antinode. To test the viability of this proposal, we study a model system with a similar flat dispersion in the absence of disorder.

\begin{figure} 
	\captionsetup{width=1\columnwidth, justification = RaggedRight}
	\includegraphics[width=1\columnwidth]{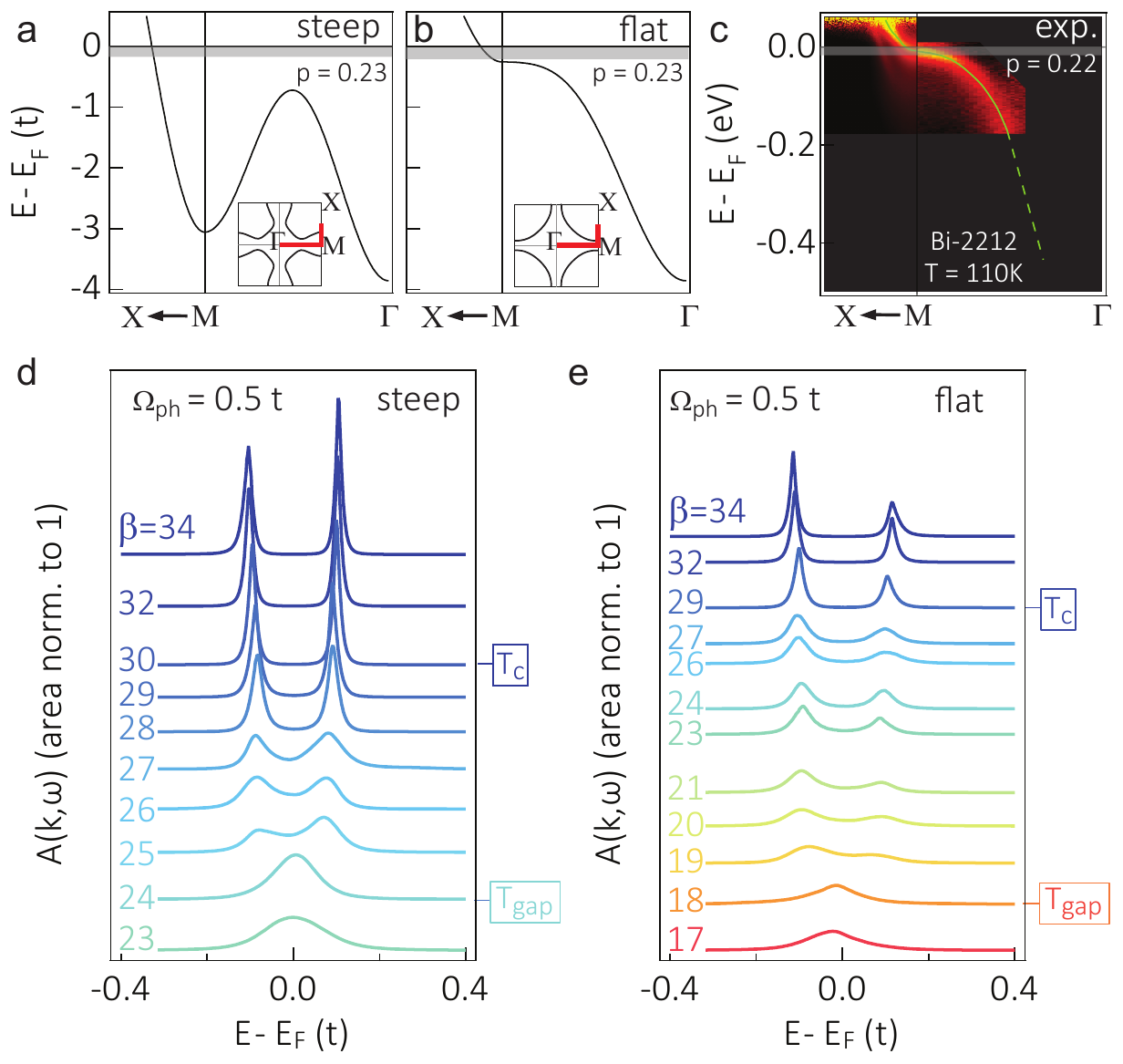}
	\caption{Band structure contribution to the superconducting fluctuation examined by quantum Monte Carlo simulations. Tight-binding band structures with (a) a steep dispersion and (b) a shallow and flat dispersion near ($\pi$,0). The along high symmetry cuts are shown by red lines on their respective Fermi surfaces in the insets. (c) Experimentally measured normal state spectra along the same high symmetry cuts for $p$ = 0.22 samples. Grey bars are the superconducting gap sizes. Calculated temperature dependent single particle spectra at the antinode for (d) the steep dispersion and (e) the shallow flat dispersion, when anisotropic superconductivity is induced by a $B_{1g}$ phonon. Spectral intensity is normalized so the total area is 1. The curves are offset proportional to temperature $T = 1/\beta$. The electron-phonon coupling strength $\lambda$ is set to 0.25.}
	\label{fig:figure4}
\end{figure}

It is well known that a flat band comes with a large effective mass, hence low superfluid density or phase stiffness $\sim\frac{n}{m^*}$. However, for overdoped cuprates, the band dispersion is only flat near the antinode. This raises the interesting question of \textit{what the effect on superfluid density will be if flat band dispersion only exists in part of the Brillouin zone}. We address this question with a sign-problem free quantum Monte-Carlo simulation on similar band structure as the overdoped cuprates.

To get rid of the fermion sign problem, and still maintain the strongest pairing interaction at the antinode, we choose a model where the band structure is tuned to mimic that of an overdoped cuprate, while the pairing interaction is mediated by a \textit{d}-form factor ($B_{1g}$) Einstein phonon coupled at $\lambda = 0.25$. \footnote{Note that we do not intend this to be a realistic model for the cuprates. However, the question to address, namely, what impedes the superconducting phase coherence when the band structure contains flat band in part of the Brillouin zone, remains valid. Thus the simulation below aims to address a matter-of-principle question, rather than a quantitative description of the cuprate material.} We compare the quantum Monte Carlo results for band structures without and with a flat antinodal dispersion. The chosen dispersions (near ($\pi$,0)) are shown in Fig.~\ref{fig:figure4}(a)(b) (experimental band structure is shown in Fig.~\ref{fig:figure4}(c)). In both cases, the system exhibits highly anisotropic $s$-wave superconducting ground states with the gap maximum at ($\pi$,0).

\begin{figure} 
	\captionsetup{width=1\columnwidth, justification = RaggedRight}
	\includegraphics[width=1\columnwidth]{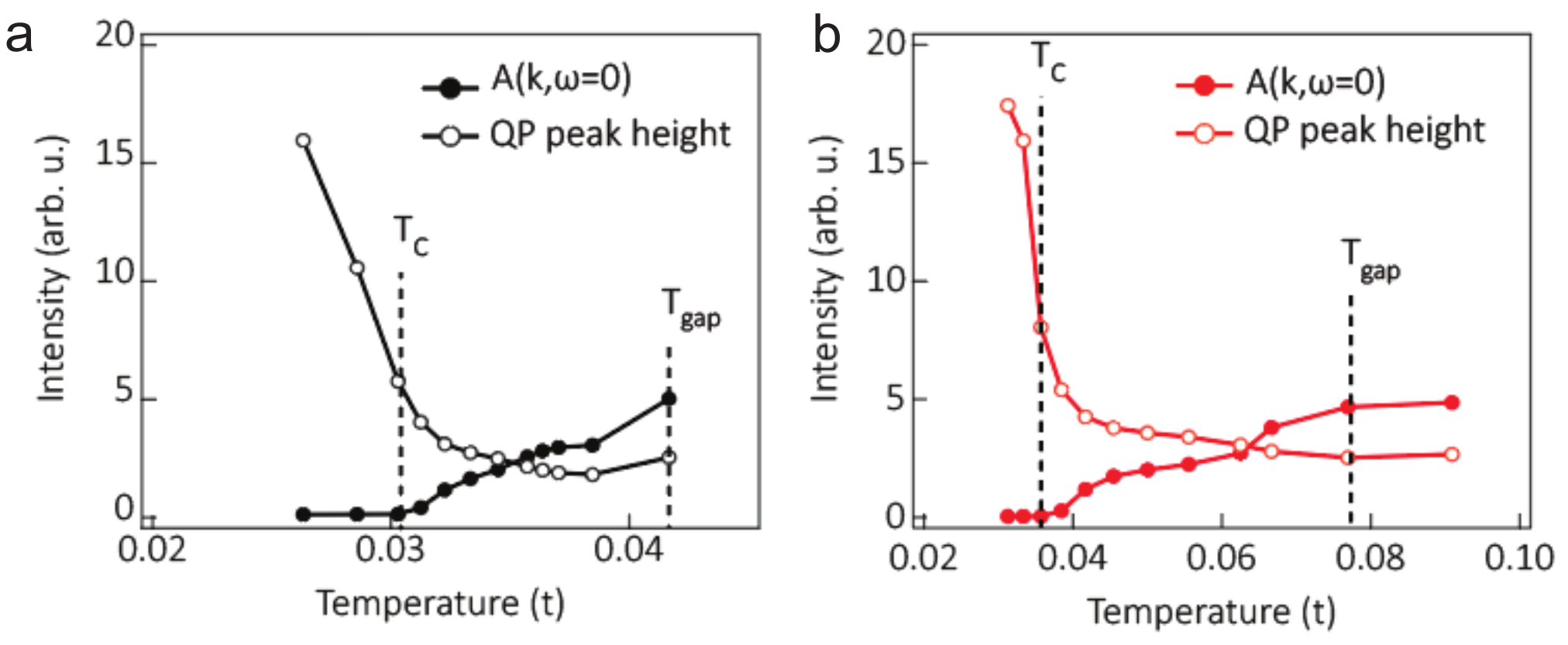}
	\caption{Temperature dependent spectral evolution from simulation. Zero energy spectral weight (solid circle) and quasiparticle peak width (open circle) for (a) the steep dispersion and (b) the shallow and flat dispersion near ($\pi$,0).}
	\label{fig:figure5}
\end{figure}

Figure~\ref{fig:figure4}(d) and (e) plot temperature dependent EDCs near the antinode for band structures shown in (a) and (b) respectively.~\footnote{The particle-hole asymmetry is due to the small momentum deviation from the exact $\mathbf{k}_F$ in the finite lattice used for the simulation.} The superconducting gap opening temperature $T_\text{gap}$, identified by the closure of the antinodal spectral gap, is 33\% higher in the case where the band structure is flat in the antinodal region. This is because the pairing interaction mediated by the $B_{1g}$ phonon can take advantage of the large density of states where flat band exists. On the other hand, the Kosterlitz-Thouless $T_c$ for a band structure with a flat antinodal dispersion is only 3\% higher (see Fig.~S\ref{fig:figureS70} for $T_c$ determination), leaving a much wider temperature window where the gap has opened but phase coherence has yet to be established. Consistent with the enhanced thermal fluctuations above $T_c$, the zero temperature superfluid density - the quantity central to previous magnetic and optical investigations~\cite{bovzovic2016dependence,mahmood2019locating} - is 45\% lower with the flat dispersion with quantum fluctuations considered~\cite{li2020inprep}. These results suggest that even without disorder, the flat dispersion near the antinode plays an important role in suppressing the superfluid density (or the normal state Drude weight) in clean systems. Additionally, qualitative consistency is found between the experiment and simulation derived spectral lineshape evolution from the normal state to zero temperature. Shown in Fig.~\ref{fig:figure5} (a) and (b) are the temperature evolutions of the zero energy spectral weight (solid circle) and the finite energy quasiparticle peak height (open circle) for the steep band (black) and partially flat band (red) respectively, based on the analysis used in Fig.~\ref{fig:figure3}. Indeed, the spectral weight on $E_F$ starts to deplete at $T_\textrm{gap}$, and the rapid growth of quasiparticle peak height coincides with the onset of the global phase coherence at $T_c$. While more microscopic theories are clearly needed to understand how global phase coherence can imprint itself on the single-particle spectral properties~\cite{carlson2000dimensional}, our work adds numerical and momentum-resolved insights to a growing number of similar observations in other superconducting systems~\cite{feng2000signature,sacepe2011localization,cho2019strongly}.

\section{Discussions}

Real systems always contain disorder, which is also expected to suppress the superfluid density via the breaking of Galilean invariance~\cite{leggett2006quantum}. In this regard, the flat band dispersion amplifies the effect of disorder via increased scattering between antinodes, which is pair breaking due to the \textit{d}-wave symmetry~\cite{he2014fermi}. Recent low temperature STS studies in heavily overdoped Bi$_2$Sr$_2$Ca$_2$Cu$_3$O$_{10+\delta}$ also indicate presence of such strong ($\pi$,$\pi$) scattering~\cite{zou2021particle}.\footnote{In this case, particle-hole symmetry of the superconducting quasiparticles is locally broken due to ($\pi$,$\pi$)-scattering induced level splitting on the order of 2-3~meV, which will spatially average to particle-hole symmetric quasiparticle peaks on length scales longer than a few in-plane lattice constants.} Hence the flat dispersion and disorder can have a combined role in driving the superconductor to metal transition in the heavily hole-doped cuprates~\cite{li2020inprep}. The answer to which effect is dominant will likely vary between different families of compounds. To verify this proposal further, one may consider direct measurements of phonon or magnetic scattering in the ($\pi$,$\pi$) channel, or quasiparticle interference via STS, or tuning the van Hove point in different overdoped cuprate compounds.

Our results also emphasize the impact of the low-energy electronic anisotropy on the transport and superconducting properties. While it has become customary to use the Fermi energy $E_F$ as a liaison to describe the carrier density and Drude weight in metallic systems, non-parabolic and anisotropic low-energy electronic structures can cause superconducting fluctuations of dramatically different strengths in systems with identical $E_F$. As such, plain ratios such as $T_c/T_F$ or $\Delta/E_F$ should be used with caution to categorize a material's superconducting properties, especially in the absence of detailed knowledge of its low-energy electronic structures. In heavily hole-doped cuprate superconductors, despite vastly different disorder levels across different families, the ubiquitously shallow van Hove point and its coincidence with the maximum \textit{d}-wave pairing gap may be one common thread underlying the cuprates' much stronger superconducting fluctuations compared to other bulk superconductors (Fig.~\ref{fig:figure0}).

In summary, we show in this work that despite the much improved three dimensionality and metallicity in a heavily overdoped cuprate, not only is there a persistent spectral gap, but it is also of purely superconducting nature rather than a continuation of the incoherent pseudogap from the underdoped cuprates. We find that a shallow, flat van Hove singularity can exacerbate the destruction of phase coherence by itself or together with disorder. We also provide experimental and numerical connections between the global phase coherence and the single particle spectral function. Given the generality of our model, it is plausible that the cooperative effects of flat band and disorder can play a role in the fluctuating phenomena in other systems, such as twisted graphene systems~\cite{liu2020tunable} and the nickelates~\cite{li2019superconductivity,choi2020quantum}. In practice, our results provide not only clear band structure targets to engineer the phase stiffness, but also the basis for $T_c$ enhancement at the interface between superconductors with strong pairing and those with large phase stiffness~\cite{kivelson2002making}.

\section*{Acknowledgement}
The authors wish to thank Steve Kivelson, Edwin Huang, Yao Wang, Junfeng He, Douglas Scalapino, Yoni Schattner, Jiecheng Zhang, Erik Kountz, Rudi Hackl, Peter Hirschfeld and Jan Zaanen for helpful discussions. The works at Stanford University and Stanford Synchrotron Radiation Lightsource, SLAC National Accelerator Laboratory, are supported by the U.S. Department of Energy, Office of Science, Office of Basic Energy Sciences under Contract No. DE-AC02-76SF00515. Part of this work is performed at the Stanford Nano Shared Facilities (SNSF), supported by the National Science Foundation under award ECCS-1542152. The works at the University of California, Berkeley and Lawrence Berkeley National Laboratory are supported by U.S. Department of Energy, Office of Science, Office of Basic Energy Sciences, Materials Sciences and Engineering Division under Contract No. DE-AC02-05-CH11231 within the Quantum Materials Program (KC2202). The computational part of this research is supported by the U.S. Department of Energy, Office of Science, Office of Advanced Scientific Computing Research, Scientific Discovery through Advanced Computing (SciDAC) program. TW and XHC acknowledge the support from the National Key R\&D Program of the MOST (Grant No. 2017YFA0303001) and the National Natural Science Foundation of China (Grants No. 11888101). DHL acknowledges support from the Gordon and Betty Moore Foundation's EPiQS initiative Grant GBMF4545. YH acknowledges support from the Miller Institute for Basic Research in Science.

\bibliographystyle{apsrev4-1}
%

\newpage
\renewcommand\thefigure{\arabic{figure}}
\setcounter{figure}{0} 
\setcounter{section}{0} 

\section{Material preparation}

\begin{figure*}
	\captionsetup{width=2\columnwidth,justification=RaggedRight}
	\includegraphics[width=1.5\columnwidth]{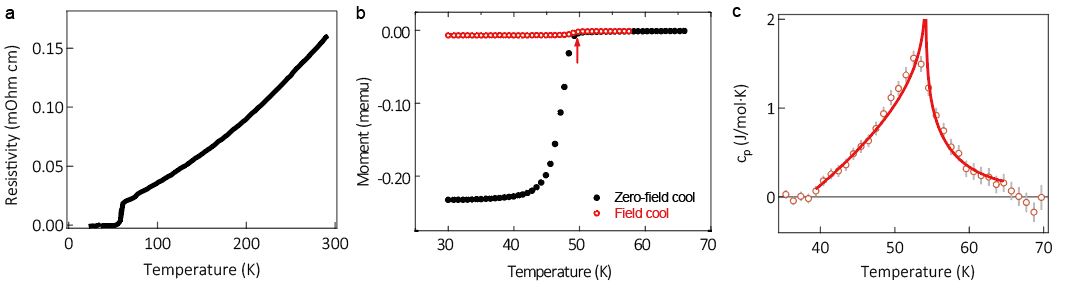}
	\caption{Bulk characterization of $p$ = 0.24 sample. (a) In-plane resistivity. (b) Magnetic susceptibility at 5 Oe with field parallel to the c-axis. (c) Heat capacity with background subtracted around the superconducting transition. Red solid line is a fit to the 3D-XY divergence.}
	\label{fig:figureS0}
\end{figure*}

\begin{figure}
	\captionsetup{width=1\columnwidth,justification=RaggedRight}
	\includegraphics[width=1\columnwidth]{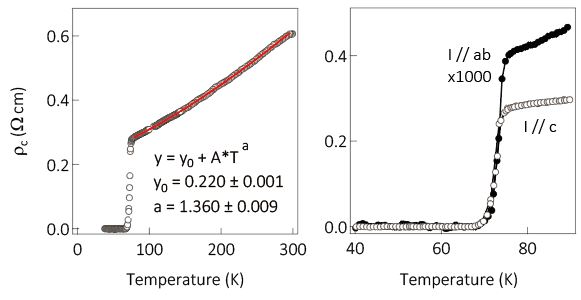}
	\caption{Resistivity of $p$ = 0.22 Bi-2212. (a) c-axis resistivity and the power law fitting. (b) Both in-plane and c-axis resistivity close to the superconducting transition. Note that the resistive zero temperature corresponds to the onset temperature of the Meissner diamagnetic signal in Fig.~1(a).}
	\label{fig:figureS6}
\end{figure}

The Bi-2212 samples are grown with the optical floating zone method with (p = 0.22) and without (p = 0.22, 0.23 and 0.24) Pb doping. The purpose of the Pb doping is to suppress a superstructure caused by the Bi-O layer bonding buckling along the diagonal Cu-Cu direction. Up to 400~bar oxygen annealing at 500 degree Celsius for 48 hours is used to press excess oxygen into the material to reach the extreme doping at p = 0.23-0.24. The physical property characterization of our p = 0.24 sample is shown in Fig.~\ref{fig:figureS0}. The crystal orientation is then determined from back scattering Laue prior to ARPES experiments. Apart from the reduced $T_c$, the effect of superoxygenation can also be seen in the reduced resistive anisotropy between the $c$-axis and $ab$-plane as shown in Fig.~S\ref{fig:figureS6}, a 2-decade reduction comparing to the optimally doped case~\cite{chen1998anisotropic}.

\section{NMR measurements}

\begin{figure}
	\captionsetup{width=1\columnwidth,justification=RaggedRight}
	\includegraphics[width=0.8\columnwidth]{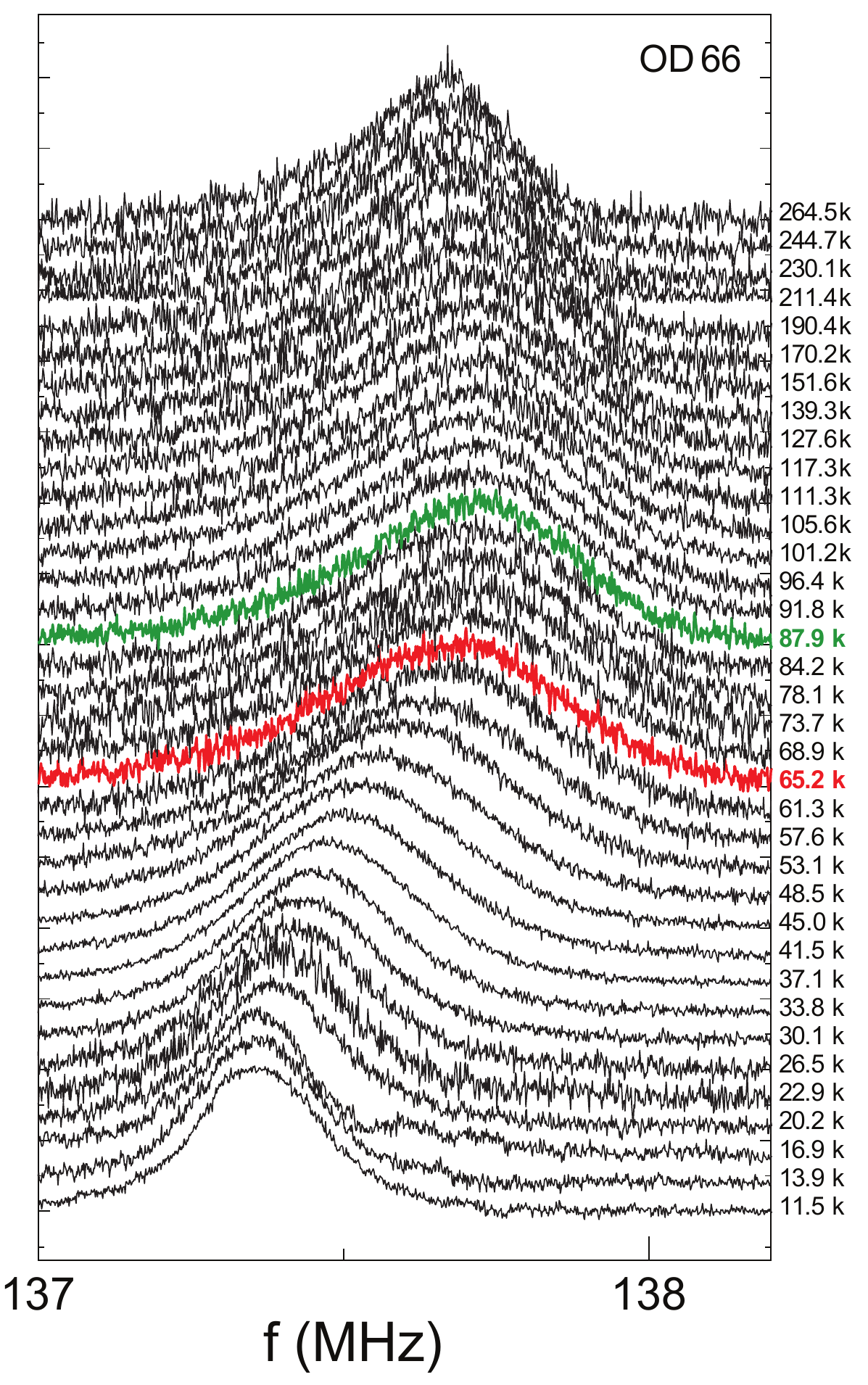}
	\caption{Raw data for the nuclear magnetic resonance measurement in OD66 ($p$ = 0.22) sample. Red and green lines represent superconducting $T_c$ and $T_\text{pair}$ respectively.}
	\label{fig:figureS130}
\end{figure}

\begin{figure}
	\captionsetup{width=1\columnwidth,justification=RaggedRight}
	\includegraphics[width=0.8\columnwidth]{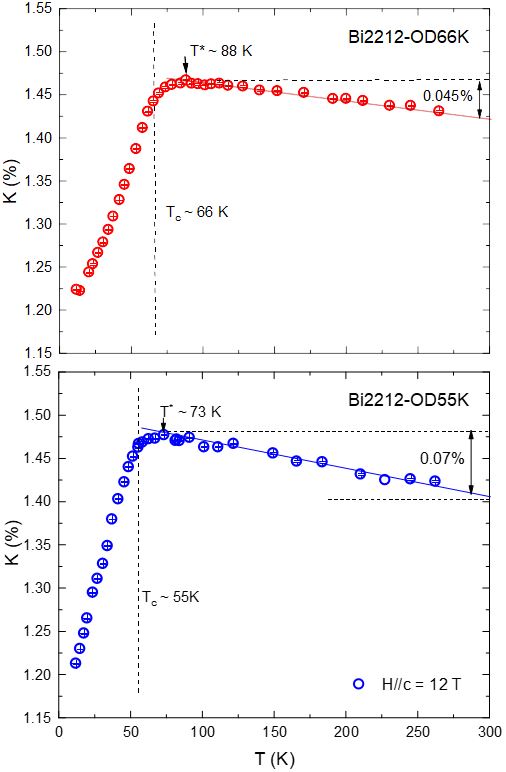}
	\caption{Temperature dependent Knight shift of $^{63}$Cu in Bi-2212 of doping (top) $p$ = 0.22 and (bottom) $p$ = 0.23. Note the stronger low-temperature uplift (ferromagnetic response) in $p$ = 0.23 sample.}
	\label{fig:figureS8}
\end{figure}

The $^{63}$Cu (I = $\frac{3}{2}$) NMR spectra for the central transition ($\frac{1}{2} \rightarrow -\frac{1}{2}$) are obtained by scanning RF frequency and summing spin-echo intensity under an external magnetic field of 12 Tesla parallel to the $c$ axis. The nuclear gyromagnetic ratio used for $^{63}$Cu is $\gamma$ = 11.285~MHz/T. The NMR coils were made by copper wires. The $^{27}$Al ($\gamma$ = 11.0943 MHz/T) NMR signal from 0.0008~mm-thick aluminum foils (99.1\% purity) was used to calibrate the external field at the sample position. For the $^{63}$Cu NMR spectrum measurements, a Hahn-echo pulse sequence of $\frac{pi}{2}-\tau-\pi-\tau-$echo was employed, with a pulse length $t_w(\frac{pi}{2})$ = 3~$\mu s$ and a delay $\tau$ = 25~$\mu s$. The Knight shift extracted from the central spectrum with $H \parallel c$ is purely magnetic shift without the quadrupole contribution. In general, the Knight shift (K) probes the uniform local spin susceptibility. The resonance frequency of the central spectrum with $H \parallel c$ can be expressed as: $f_\text{res}=(1+K) f_0$, where $f_0=\gamma H$ is the Larmor frequency of a bare nucleus with gyromagnetic ratio $\gamma$ under a magnetic field $H$, and $f_\text{res}$ is the observed resonance frequency in real materials.

Figure~S\ref{fig:figureS130} shows the raw temperature dependent resonance spectra of a $p$ = 0.22 sample. The high statistics shows unambiguous reduction of the resonant peak centroid from $T_\text{pair}$ (green) to $T_c$ (red). Figure~S\ref{fig:figureS8} shows the Knight shift in both the $p$ = 0.22 and 0.23 samples. Upon cooling, a clear upshift in frequency is seen, indicating ferromagnetic correlations. Moreover, such an upshift is more pronounced in the $p$ = 0.23 sample, suggesting potentially increasing ferromagnetic fluctuations as the van Hove point is approached.

\section{Details of quantum Monte Carlo simulation}

\begin{figure*}
	\captionsetup{width=1.8\columnwidth,justification=RaggedRight}
	\includegraphics[width=1.8\columnwidth]{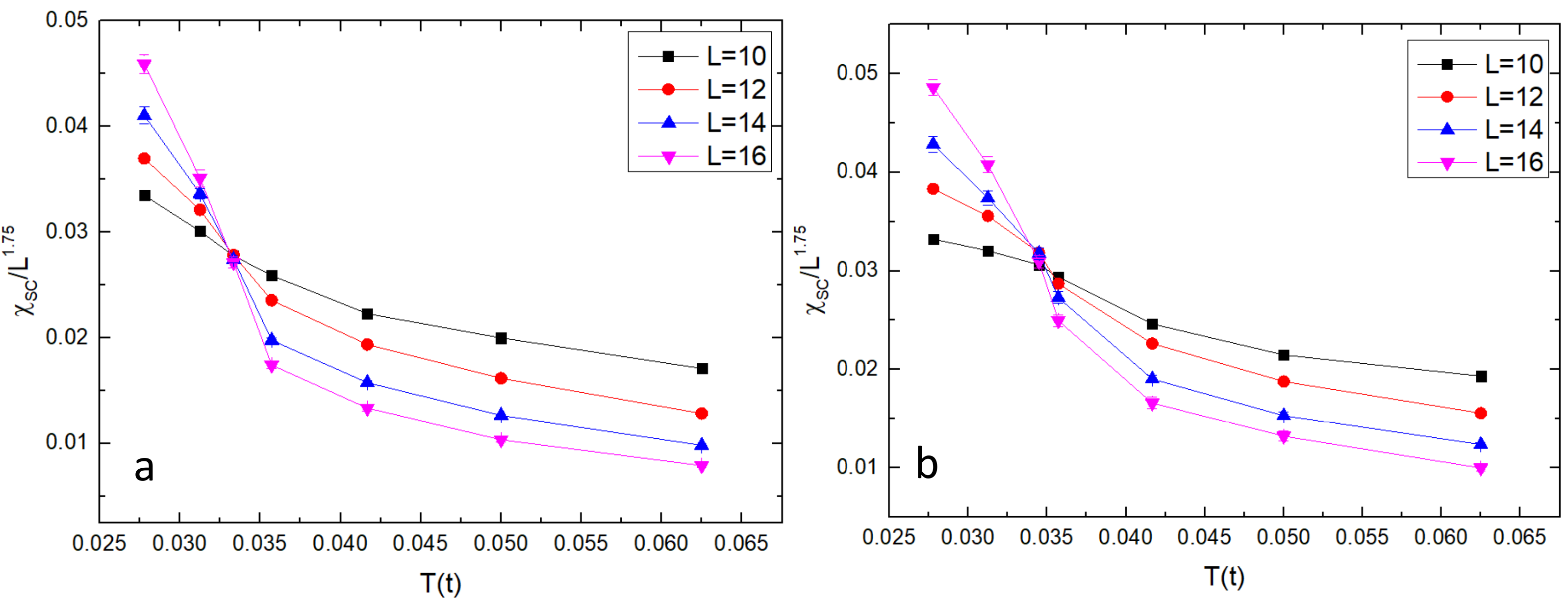}
	\caption{The scaled SC susceptibilities $\chi_{sc} L^{-1.75}$ as a function of temperature for linear system size $L=10,12,14,16$. The crossing point of the curves for different system sizes marks the SC transition temperature $T_c$. (a) Regular band structure: $T_c \approx \frac{t}{30}$.  (b) The partly flat band with shallow and flat dispersion in antinodal region: $T_c \approx \frac{t}{29}$.}
	\label{fig:figureS70}
\end{figure*}

 To examine the effects of flat band dispersion near the antinode  on suppressing the zero temperature superfluid density and enhancing the SC fluctuations, we construct two different band structures distinguished by  i) the absence and ii) the presence of flat band dispersion near the $(\pi,0)$ and $(0,\pi)$ points in the Brillouin zone. In case (ii) the band structure is constructed to mimic that of an overdoped cuprate. The Fermi surface and band dispersions along high symmetry cuts of these models are shown in the main text. 
 
We perform sign problem free determinant quantum Monte Carlo simulation to investigate the superconductivity triggered by the coupling between electrons and a $B_{1g}$ phonon in these different band structures. The Hamiltonian used in our calculation is given by:
\begin{equation}
\label{model}
\begin{split}
H =& ~H_e + H_p + H_{ep} \nonumber\\
H_e =& - \sum_{ij\sigma} t_{ij} (c^\dagger_{i\sigma} c_{j\sigma} + h.c.) - \mu \sum_i \hat{n}_{i\sigma} \nonumber\\
H_p =& \sum_i \frac{\hat{P}^2_i}{2M} + \frac{K}{2} \hat{X}_i^2 \nonumber\\
H_{ep} =& ~g \sum_{i\sigma} \hat{X}_i (c^\dagger_{i\sigma} c_{i+x\sigma} + c^\dagger_{i\sigma}c_{i-x\sigma}\\
& \- c^\dagger_{i\sigma} c_{i+y\sigma} - c^\dagger_{i\sigma}c_{i-y\sigma} + h.c.)
\end{split}
\end{equation}

\noindent Here $c_{i\sigma}$ annihilates an electron on site $i$ with spin polarization $\sigma = \uparrow/\downarrow$, $\mu$ is the chemical potential and $\hat{n}_{i\sigma}$ is the electron number operator. $t_{ij}$ is the electron hopping integral between site $i$ and $j$ which is truncated beyond third neighbor. The choices for these parameters are as follows. i) For the flat dispersion free model $t=1$, $t'=-0.4$, $t''=0.68$, $\mu = -1.263$. ii) For the model with flat dispersion $t=1$, $t'=-0.05$,$t''=0.2$, $\mu = -0.778$. These parameters are chosen such that the doping level (23\%) and the global Fermi energy (3.857$t$) are the same for the two different band structures. Thus, the most important difference between the band structures i) and ii) is the absence/presence of flat dispersion along the Brillouin boundary.

In this model, $H_p$ describes an Einstein phonon with frequency $\omega = \sqrt{\frac{K}{M}}$, and $\hat{X}_i$/$\hat{P}_i$ is the associated displacement/momentum operator. The term $H_{\rm ep}$ in model describes the modulation of the electron nearest neighbor hopping by the phonon, where $g$ is the coupling constant. It is worthwhile noting that such modulation occurs with an opposite phase for the x/y direction bonds. This $d$-wave form factor is characteristic of the coupling to the $B_{1g}$ phonon. To quantify the strength of the electron-phonon interaction we introduce the dimensionless coupling constant $\lambda = [2\sum_{kq} g(k,q)^2 \delta(\epsilon_k)\delta(\epsilon_{k-q})]/N K \sum_k \delta(\epsilon_k)$, where $g(k,q)$ is the electron-phonon coupling constant in momentum space, $N$ is the number of momentum points in our finite-size system, and  $\epsilon_k$ is the electron dispersion. In the actual calculation $\lambda$ is set to 0.25 and the phonon frequency $\omega$ is set to 0.5$t$. This value of $\omega$ is larger than that of the $B_{1g}$ phonon in the cuprate. The reason we did not choose a smaller $\omega$ is because that makes it more difficult to extract the superconducting gap accurately in our finite size system ($L_{\rm max}=16$).  Moreover, a larger $\omega$ reduces the auto-correlation time and significantly speeds up the simulation. Importantly, because the same phonon frequency is used in conjunction with both band structures, we do not expect that its larger value  will jeopardize our main goal, namely, studying the difference in the zero temperature superfluid density and the degree of superconducting fluctuations resulting from the difference in band structures.

\subsection{The phase coherence and gap opening temperature}
To determine the superconducting (SC) phase coherence temperature, we evaluate the SC susceptibility via $\chi_{SC} = \int_0^\beta \langle\Delta(\tau)\Delta^\dagger(0)\rangle$, where $\Delta$ is the SC order parameter. The SC phase coherence temperature  $T_c$ (or  the Kosterlitz-Thouless transition temperature) is determined through the well-known finite size scaling behavior by plotting $\chi_{SC} L^{-1.75}$ as a function of temperature for different system sizes $L$. The crossing point of different curves marks the K-T transition temperature $T_c$. This is shown in Fig.~S\ref{fig:figureS70}. Note that the pairing symmetry caused by the electron-phonon interaction is \textit{s}-wave.

To determine the gap opening temperature we compute the electron spectral function at the Fermi momentum in the antinodal region. This is achieved by first computing the imaginary-time Green's function $G(k,\tau) = \int_{-\infty}^{\infty} d\omega \frac{e^{-\omega(\tau-\beta/2)}}{2\cosh(\beta\omega/2)} A(k,\omega)$, then using the stochastic analytical continuation to extract the $ A(k,\omega)$ ~\cite{Sandvik1998}. The results are shown in the main text. Below the SC transition temperature $T_c$, $A(k,\omega)$ exhibits two sharp particle-hole symmetric peaks. Above $T_c$, these peaks broaden but remain well-defined in a wide temperature window, the energy separation between them defines the normal state gap. Eventually, at sufficiently high temperature, the spectral gap is no longer discernible as $A(k,\omega)$ shows a single broad peak. From such a spectral function evolution we extract the gap opening temperature $T_{\textrm{gap}}$. The main result of the Monte-Carlo simulation is the fact that $T_{\textrm{gap}}/T_c$ is significantly larger for the band structure involving a flat dispersion at where the superconducting gap maximizes.

\subsection{The zero-temperature superfluid density}
The zero-temperature superfluid phase stiffness is determined by computing the current-current correlation: $\rho_{sc} = -\langle {\rm KE}_x\rangle - \Lambda_{xx}(q_x=0, q_y\rightarrow0)$, where ${\rm KE}_x$ is kinetic-energy density associated with an x-oriented bond and $\Lambda_{xx}(\vec{q})= \sum_i \int_0^\beta e^{-i \vec{q}\cdot\vec{r}_i} \langle J_x(r_i,\tau) J_x(0,0)\rangle$. In the projector QMC, we evaluate the ground-state
expectation values of an observable according to $\langle \hat{O} \rangle  = \lim_{\theta\rightarrow \infty} \langle \psi_T\mid e^{-\theta H } \hat{O} e^{-\theta H} \mid\psi_T \rangle/ \langle \psi_T\mid e^{-2\theta H}\mid \psi_T \rangle$, where $\theta$ is projective parameter and $\mid \psi_T \rangle$ is the trial wave function. In the calculation, we use $\theta = 100/t$ and have checked that this value is sufficiently large to ensure  convergence. Due to the long autocorrelation time of quantum Monte Carlo at zero temperature, we run a 1000 independent Markov chains with 10000 space-time  sweeps in the simulation, which is rather resource heavy. The results show that the superfluid phase stiffness is $$\rho_{\rm s,~flat~band}=0.126 \pm 0.004$$ in the model with flat band dispersion, and $$\rho_{\rm s,~ no~flat~band}=0.231 \pm 0.007$$ in the model without a flat dispersion. Therefore $${\rho_{\rm s,~flat~band}\over \rho_{\rm s,~ no~flat~band}}\approx 0.55.$$

\section{ARPES measurements}

\begin{figure}
	\captionsetup{width=1\columnwidth,justification=RaggedRight}
	\includegraphics[width=1\columnwidth]{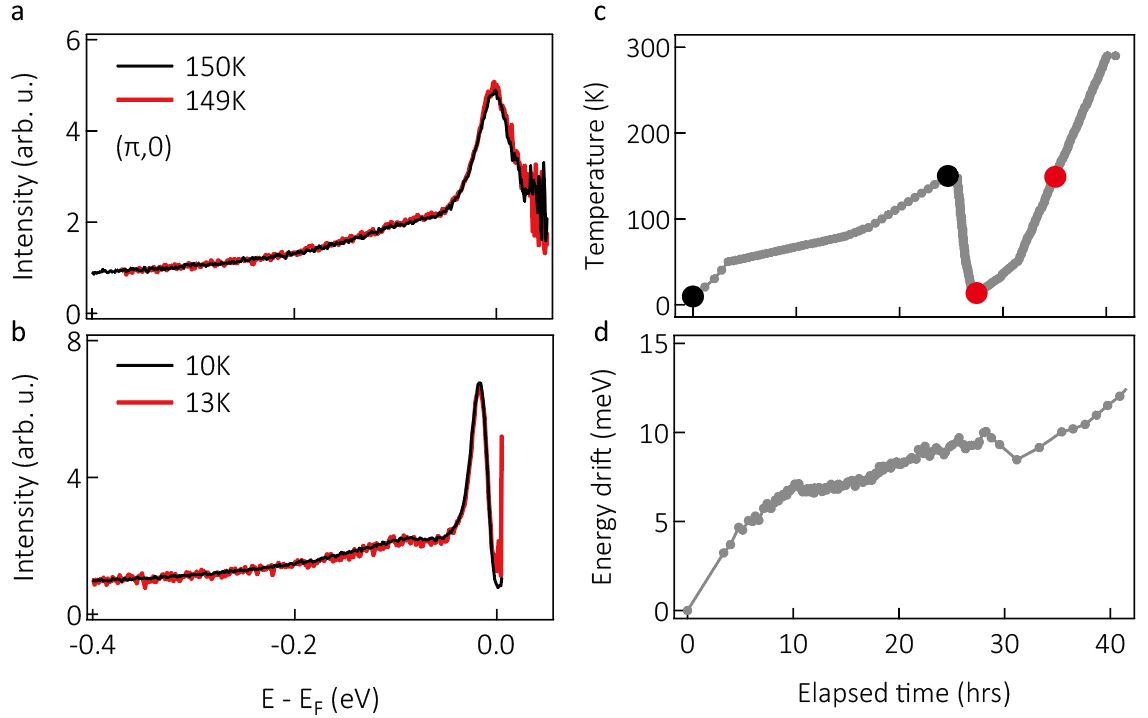}
	\caption{Typical ($\pi$,0) Fermi function divided EDCs before and after slow temperature cycles. (a) Temperature cycles before and after around 150~K. (b) Temperature cycles before and after around 10~K. (c) Temperature history of the measurements that produced (a) and (b). (d) Relative photon energy drift history over the course of the measurement on the $p$ = 0.22 sample. Sample spectra collected between two closely monitored reference gold spectra are corrected according to the straddled Fermi level between the gold data.}
	\label{fig:figureS2}
\end{figure}

\begin{figure}
	\captionsetup{width=1\columnwidth,justification=RaggedRight}
	\includegraphics[width=1\columnwidth]{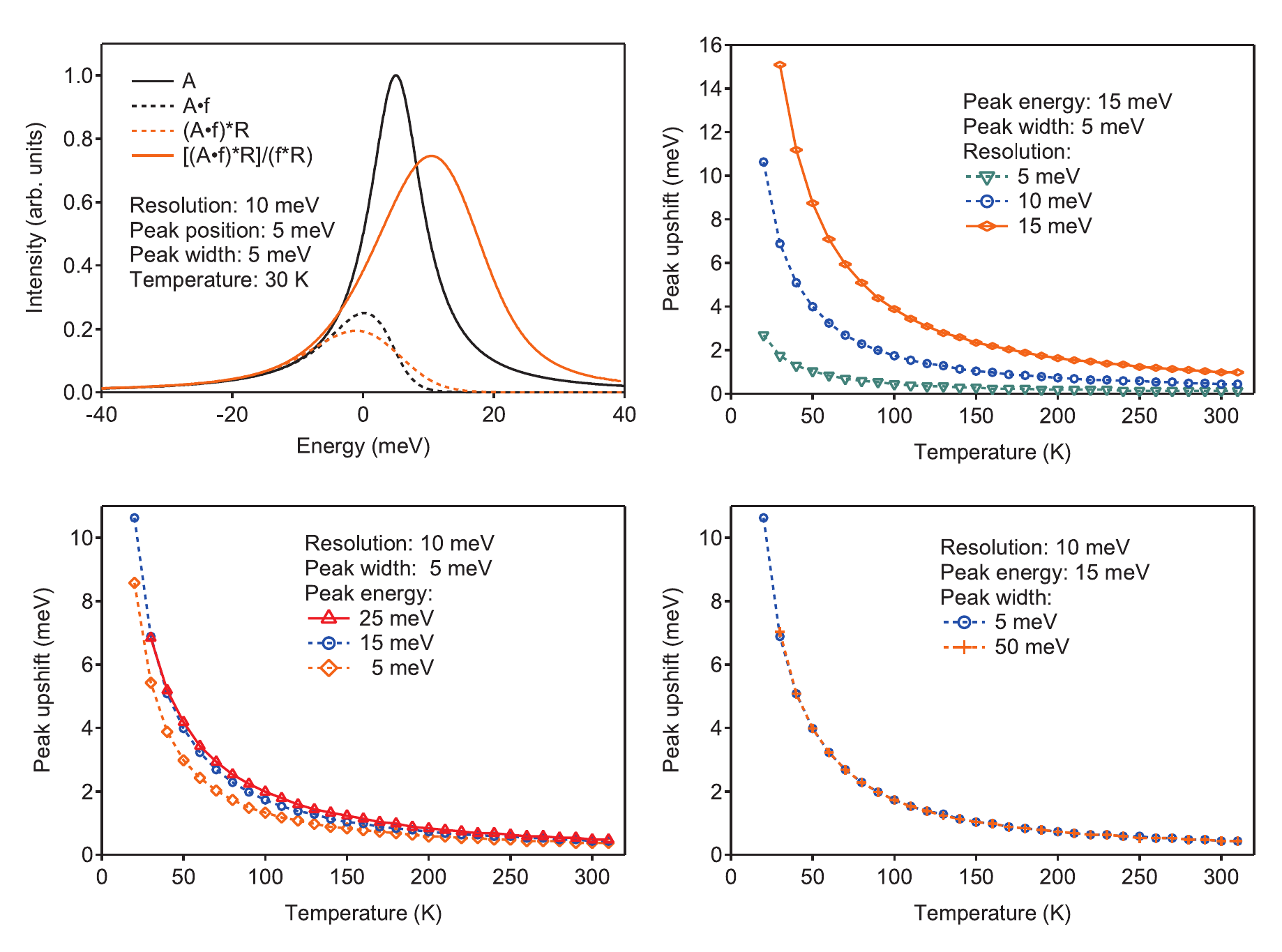}
	\caption{Resolution effect in the restoration of states above $E_F$ via Fermi-function division. Top left: At the given temperature of 30~K, a simple Lorenztian peak (A, solid black line) becomes thermally populated (black shaded line) after product with a Fermi function (f). It is then convolved with the instrument resolution (R) and becomes the experimentally measured line spectrum (orange dashed line). Division of this experimentally measured spectrum by the convolution of Fermi function and the instrument resolution results in the restored spectrum (orange solid line). Due to the non-commutability of the division and convolution process, the restored spectrum is seen to have artificially upshifted in energy comparing to the original Lorenztian. Resolution (top right), peak energy (bottom left) and peak width (bottom right) dependence of the Fermi-function division induced peak upshift.}
	\label{fig:figureS14}
\end{figure}

The photoemission measurements are performed at BL5-4 at the Stanford Synchrotron Radiation Lightsource at SLAC National Laboratory. The photon energies used are 18.4~eV (at 8~meV combined instrument resolution) and 9~eV (at 3.5~meV combined instrument resolution) with the light polarization parallel to the Cu-O bond direction. A local heater that only heats up the small thermal mass on the flip stage is used to minimize out-gassing, while the radiation shield is kept largely below 100~K throughout the experiment~\cite{chen2019incoherent}. The vacuum is kept between 1.9 and 2.5$\times$10$^{-11}$ Torr over the entire course of the experiment. Sample ageing is checked via two rounds of careful temperature cycles with ramping rate between 0.05-0.40~K/min (see Fig.~S\ref{fig:figureS2}(a)(b)(c)).

To maintain the photon energy uncertainty below 0.5~meV, we interlace the entire measurement with more than 50 rounds of polycrystalline gold measurements to provide a finely straddled incident photon energy reference at an unprecedented accuracy for synchrotron measurements (Fig.~S\ref{fig:figureS2}). To maintain high fidelity of the Fermi-function division procedure so as to obtain reliable unoccupied state information, the detector non-linearity is first corrected with respect to a calibration between total photon flux and detector count rate done on a polycrystalline gold sample. A small resolution effect resulting from the Fermi-function division process is quantitatively analyzed in Fig.~S\ref{fig:figureS14}, which yields $\sim$2-3~meV artificial spectral upshift in the present experimental condition.

\section{Persistent hole-like Fermi surface and fluctuating $d$-wave superconductivity up to $p=$ 0.24}

\begin{figure}
	\captionsetup{width=1\columnwidth,justification=RaggedRight}
	\includegraphics[width=1\columnwidth]{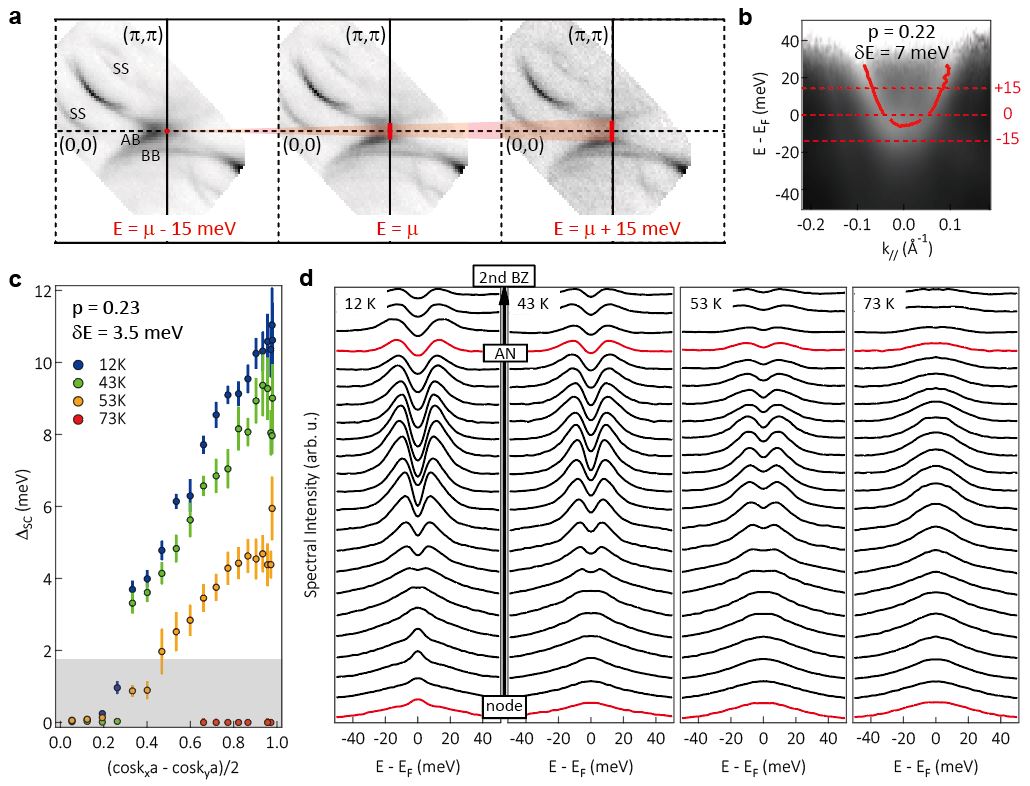}
	\caption{Normal state Fermi surface and superconducting gap anisotropy in $p$ = 0.22 and $p$ = 0.23 Bi-2212. (a) From left to right: constant energy contour of the in-plane electronic structure at 15~meV below, on, and 15~meV above the chemical potential at 104~K. SS - BiO layer superstructure. AB - antibonding band. BB - bonding band. Red bars are the antibonding band dispersion separation at the BZ boundary. (b) Energy-momentum cut along the BZ boundary cut across ($\pi$,0) taken with an energy resolution of 7~meV. Solid red line is the fitted dispersion. (c) Momentum dependence of the spectral gap along the antibonding band Fermi surface at different temperatures. Grey shade indicate half of the nominal energy resolution at 1.75~meV. (d) Symmetrized energy distribution curves (EDCs) as function of momentum at the corresponding temperatures in (c).}
	\label{fig:figureS1}
\end{figure}

Fluctuating superconductivity can be identified in even more heavily overdoped systems besides the $p$ = 0.22 system. Figure~S\ref{fig:figureS0} shows the resistivity, Meissner effect and heat capacity data in a $p$ = 0.24 sample ($T_c$ = 51~K). The Fermi surface is shown to remain hole-like (Fig.~S\ref{fig:figureS1}(a)(b)), and a $d$-wave gap uniformly closes from the node to the antinode on the entire Fermi surface when warmed up to high temperature (Fig.~S\ref{fig:figureS1}(c)(d)). Note that an intact $d$-wave gap remains at the bulk superconducting $T_c$ (orange markers in (c)).

\begin{figure}
	\captionsetup{width=1\columnwidth,justification=RaggedRight}
	\includegraphics[width=0.7\columnwidth]{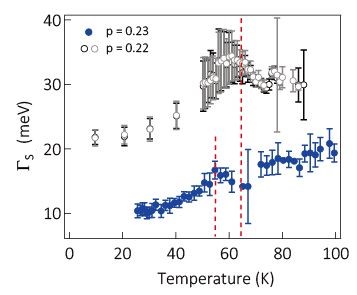}
	\caption{Fitted ``single particle scattering'' term according to the model proposed by Norman et al.~\cite{norman1998phenomenology}. Left and right red dashes denote the superconducting transition temperature in $p$ = 0.23 and $p$ = 0.22 samples respective. Black and grey open circles denote fitting from two symmetric $\mathbf{k}_F$'s on the antibonding band along the BZ boundary.}
	\label{fig:figureS4}
\end{figure}

\begin{figure}
	\captionsetup{width=1\columnwidth,justification=RaggedRight}
	\includegraphics[width=1\columnwidth]{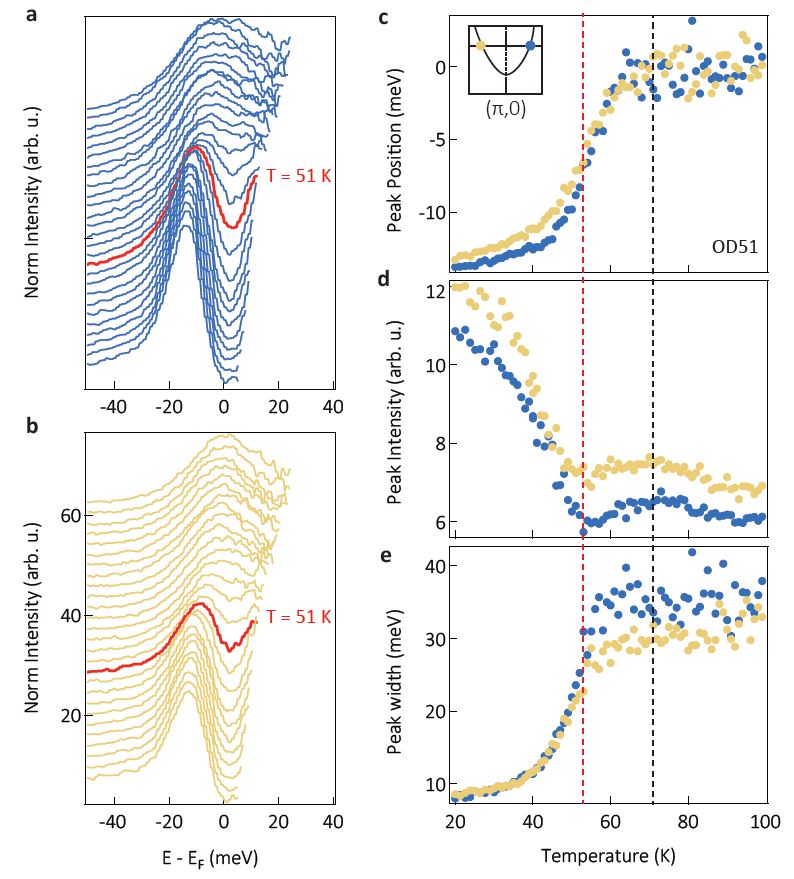}
	\caption{Temperature dependent antinodal $\mathbf{k}_F$ EDCs in $p$ = 0.24 Bi-2212. (a)(b) EDCs taken at two symmetric $\mathbf{k}_F$'s on the antibonding band with increasing temperature (from 20~K at the bottom to 98~K at the top). Peak fitting parameters for the quasiparticle - (c) position (d) height, and (e) width.}
	\label{fig:figureS7}
\end{figure}

One potential complication of using surface sensitive techniques like ARPES or STM to study Bi-2212 is the low chemical stability of the oxygen atoms on the surface layer. Sometimes this can even be used to deliberately alter the hole concentration in the system~\cite{he2018rapid,drozdov2018phase}. An intrinsic validation of the absence of sample surface ageing is the presence of two temperature scales - in particular signature for bulk $T_c$ - in the temperature dependent spectra. Other than the rapid onset of the spectral coherence as demonstrated in Fig.~3 (for $p$ = 0.22), one may also use more involved models to extract the single particle scattering rate across the superconducting transition~\cite{norman1998phenomenology}. Figure ~S\ref{fig:figureS4} shows the extracted single-particle scattering rate exhibiting an artificial singularity at the bulk $T_c$ as the quasiparticle peaks merge from two to one due to rapid loss of coherence, consistent with both the simple Lorentzian or Gaussian fitting in the main text and previous reports in Bi-2212~\cite{norman1998phenomenology}. 

Additionally, such a singularity dictated by single particle coherence formation at the bulk transition $T_c$ is also observed in even more heavily hole-doped Bi-2212 at $p$ = 0.24 (Fig.~S\ref{fig:figureS7}). It should also be noted that the clear existence of Fermi crossings at the BZ boundary directly rules out any Lifshitz transition from a hole- to electron-like Fermi surface at least till $p$ = 0.24 in Bi-2212.

\section{Gap extraction and spectral simulation in $p$ = 0.22 sample}

\begin{figure}
	\captionsetup{width=1\columnwidth,justification=RaggedRight}
	\includegraphics[width=1\columnwidth]{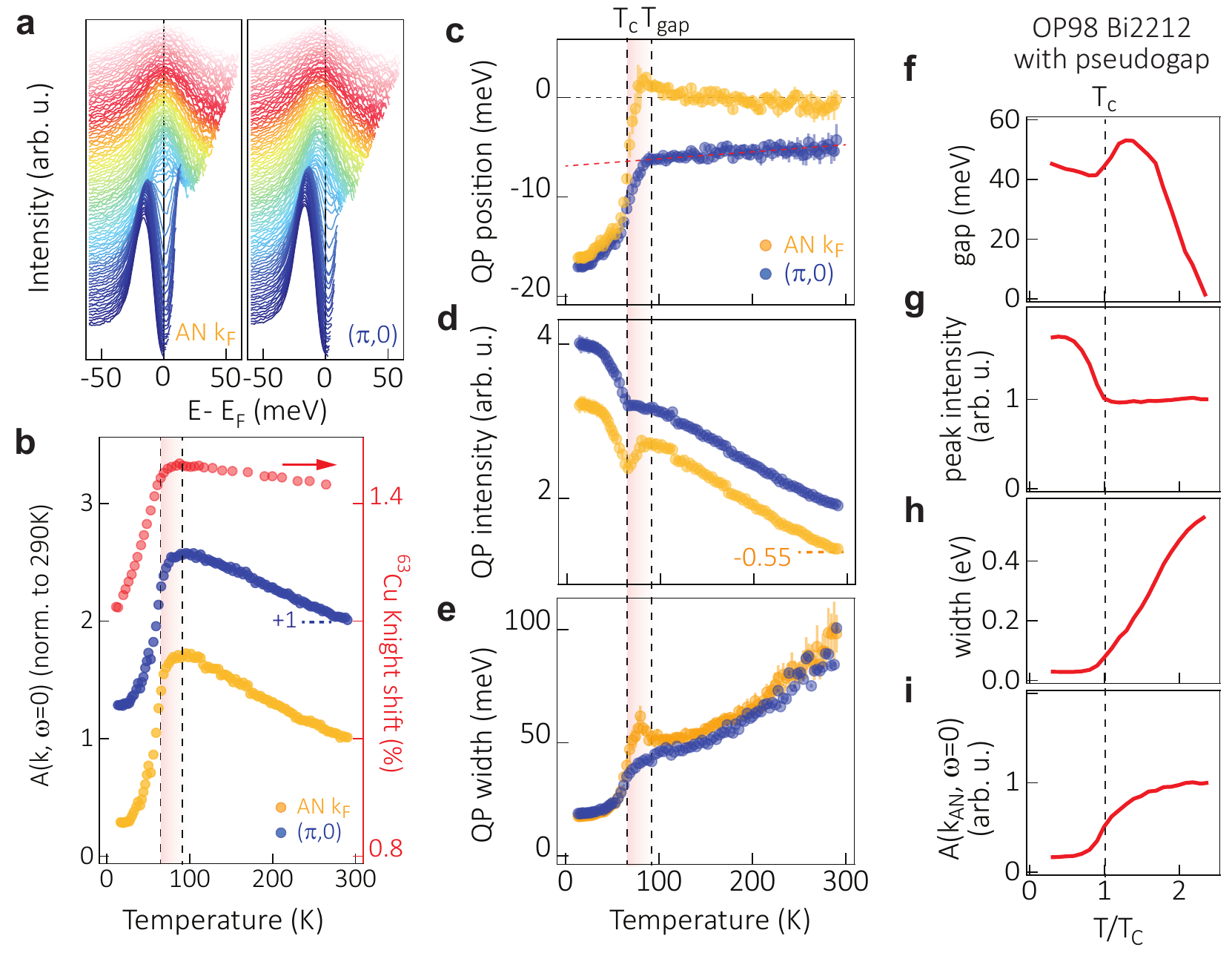}
	\caption{Full temperature dependent antinodal spectra for $p$ = 0.22 sample ($T_c$ = 66~K). (a) All temperature dependent EDCs at the antinodal $\mathbf{k}_F$ and ($\pi$,0) (10~K to 290~K, navy to pink) are normalized to the intensity at -0.4~eV binding energy. (b) Temperature dependent spectral weight at $E_F$ at two antinodal momenta from 10~K to 290~K. Red line is the NMR measured $^{63}$Cu Knight shift. (c)-(e) Single peak gaussian fitting results for the quasiparticle peak in (a) between -20~meV and +1~meV binding energy. Colored number in (d) is the offset for the curve to improve visual clarity. Antinodal spectral (f) peak energy position, (g) maximum intensity, (h) peak or hump width, (i) weight on $E_F$ in an optimally doped Bi-2212. The optimally doped Bi-2212 data is taken from Hashimoto et al.~\cite{hashimoto2015direct}.}
	\label{fig:figureS10}
\end{figure}

\begin{figure}
	\captionsetup{width=1\columnwidth,justification=RaggedRight}
	\includegraphics[width=1\columnwidth]{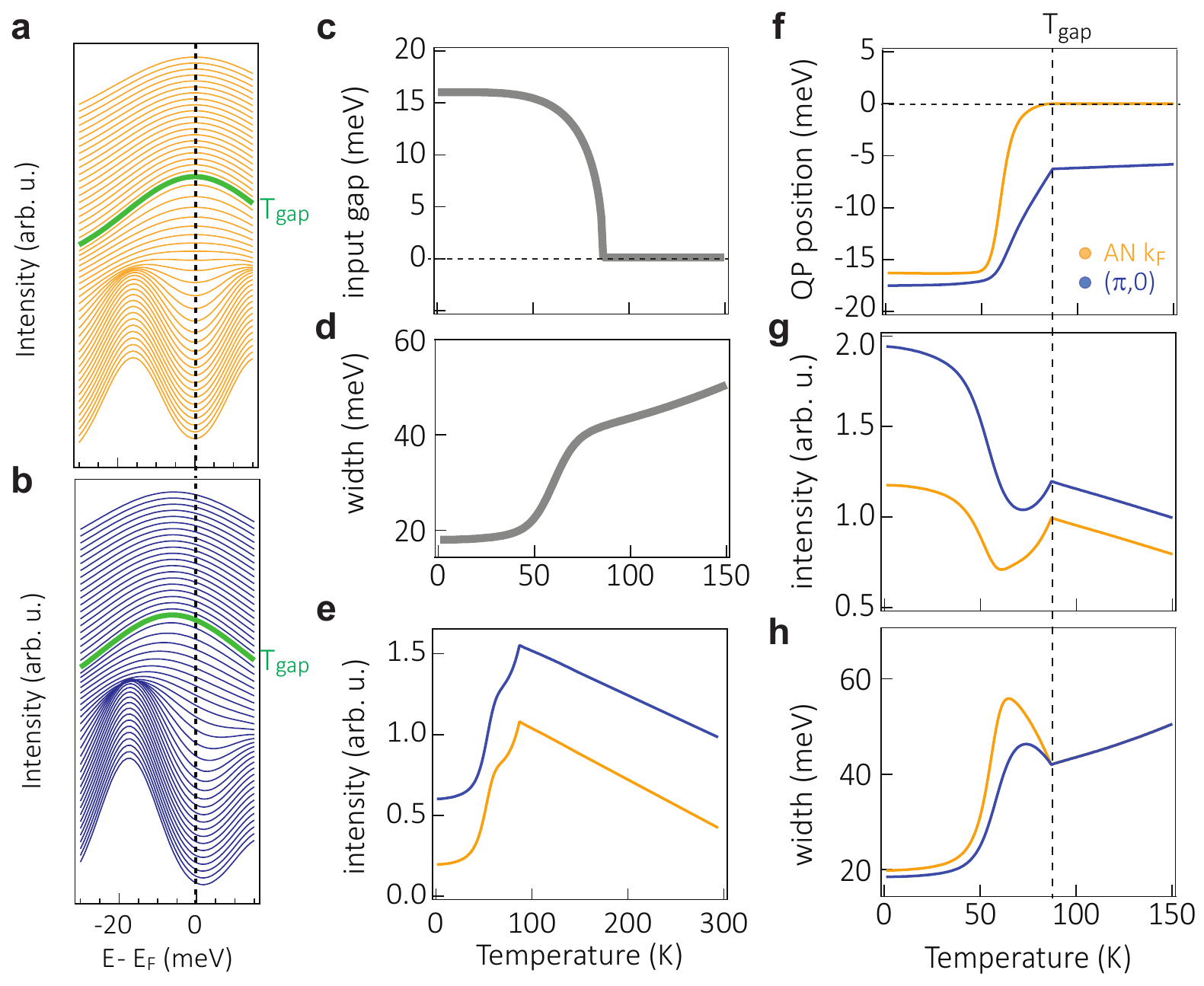}
	\caption{Simulated antinodal EDCs and fitting results. Simulated EDCs at (a) antinodal $\mathbf{k}_F$ and (b) ($\pi$,0) from 1~K (lower curve) to 150~K (upper curve). (c) BCS superconducting gap and (d) experimentally fitted quasiparticle width used as inputs for the simulation. (e) Spectral intensity at $E_F$, (f) quasiparticle peak position, (g) quasiparticle peak intensity, and (h) quasiparticle width fitted from the simulated spectra in (a) and (b).}
	\label{fig:figureS16}
\end{figure}

Fig.~\ref{fig:figureS10}(a) shows the antinodal spectra of $p$ = 0.22 sample over the full measured temperature range from 290~K to 10~K. With the wide temperature range in full display, the quasiparticle sharpening upon cooling becomes clear in both the raw Fermi divided spectra, and in the spectral intensity evolution on $E_F$ shown in Fig.~S\ref{fig:figureS10}(b). Moreover, the departure from the rising trend of the $E_F$ spectral intensity with cooling through $\sim$~100~K further highlights the fluctuating superconductivity region. With such coherent spectra, one can use very well defined single gaussian peak fit to extract the quasiparticle's position, intensity and FWHM's evolution throughout the entire temperature range measured (Fig.~S\ref{fig:figureS10}(c)-(e)). While the fitting tends to become unstable when two quasiparticle peaks merge into one especially near the antinodal $\mathbf{k}_F$ between $T_c$ and $T_\textrm{gap}$, the spectral peak near ($\pi$,0) remains at least $\sim$7~meV below $E_F$ over the entire temperature range. The particle-hole mixing ratio here is at most $\frac{E_\mathbf{k}+\epsilon_\mathbf{k}}{E_\mathbf{k}-\epsilon_\mathbf{k}}\sim\frac{16+7}{16-7}\sim23:9$, which is the maximum electron-to-hole peak intensity ratio at zero temperature. Even better, these two quasiparticle branches will be at the nearest 14~meV away from each other (at $T\sim T_\textrm{gap}$), further reducing the spectral intensity interference from the hole branch to the electron branch. Given this intrinsically asymmetric - but still substantial - particle-hole mixing, the single-peak fitting at ($\pi$,0) successfully mitigates the longstanding near-zero-gap fitting instability across $T_c$ and $T_\textrm{gap}$ in earlier cuprates gap studies. The much smoother ($\pi$,0) fitting results (blue markers in Fig.~S\ref{fig:figureS10}(c)-(e)) can be used to more reliably extract the spectral peak position, intensity and width across $T_c$. Subsequently, the peak position with low fitting error at ($\pi$,0) can provide a better measure of the superconducting gap across $T_c$ and $T_\textrm{gap}$ than the direct EDC fitting at $\mathbf{k}_F$ via the subtraction of quadrature relation: $\Delta_{(\pi,0)} = \sqrt{E_{(\pi,0)}^2-\epsilon_{(\pi,0)}^2}$ (shown in Fig.~\ref{fig:figureS130}(c)). In fact, one may use only a simple $d$-wave BCS gap (Fig.~S\ref{fig:figureS16}(c)) and the experimentally fitted quasiparticle linewidth (Fig.~S\ref{fig:figureS16}(d)) to approximate the temperature dependent EDCs at both antinodal $\mathbf{k}_F$ and ($\pi$,0) (Fig.~S\ref{fig:figureS16}(a) and (b)), via:
\begin{equation}
    \begin{split}
        A(\mathbf{k}_\textrm{F,AN},\omega) =& \frac{1}{2\sqrt{2\pi}\sigma(T)} \big(e^{-\frac{(\omega+\Delta_\textrm{AN}(T))^2}{2\sigma(T)^2}} + e^{-\frac{(\omega-\Delta_\textrm{AN}(T))^2}{2\sigma(T)^2}}\big)\\
        A(\mathbf{k}_{(\pi,0)},\omega) =& \frac{1}{\sqrt{2\pi}\sigma(T)} \bigg(\frac{E_{(\pi,0)} + \epsilon_{(\pi,0)}(T)}{2E_{(\pi,0)}}e^{-\frac{(\omega+E_{(\pi,0)}(T))^2}{2\sigma(T)^2}} \\
        &+ \frac{E_{(\pi,0)} - \epsilon_{(\pi,0)}(T)}{2E_{(\pi,0)}}e^{-\frac{(\omega-E_{(\pi,0)}(T))^2}{2\sigma(T)^2}}\bigg)\\
    \end{split}
\end{equation}
\noindent where
\begin{equation}
    \begin{split}
        \Delta(T) &\approx \Delta(0) \textrm{tanh}\bigg[1.753\sqrt{\frac{\Delta(T)^2}{\langle\Delta_\mathbf{k}(T)^2\rangle_\textrm{FS}}\big(\frac{T_c}{T}-1\big)} \bigg]\\
        E_{(\pi,0)}(T) &= \sqrt{\Delta(T)^2 + \epsilon_{(\pi,0)}(T)^2}\\
        \epsilon_{(\pi,0)}(T) &= -0.0069 + 7.07\times10^{-6}~T ~~\textrm{(eV)}\\
        \sigma(T) &= \frac{1}{1.665}\bigg[0.018 + 5.56\times10^{-7}~T^2\\
        &+ 0.010\times\Big(1+\textrm{tanh}\big(\frac{T-60}{12}\big)\Big)\bigg]~~\textrm{(eV)}
    \end{split}
\end{equation}
\noindent and $\sigma(T)$ and $\epsilon(T)$ are experimentally fitted quasiparticle linewidth and normal state van Hove point energy (red dashed line in Fig.~S\ref{fig:figureS10}(c)) respectively. The interpolated BCS gap formula is taken from Devereaux et al.~\cite{devereaux1995electronic}, where $\langle\Delta_\mathbf{k}(T)^2\rangle_\textrm{FS}\sim$~0.562~$\Delta(T)^2$ is the Fermi surface averaged $d$-wave superconducting gap squared, with the band structure taken as the tight binding band used in Fig.~\ref{fig:figureS8}(b). The zero temperature gap $\Delta(0)$ is set at 16~meV, and $T_c$ is set at $\frac{2\Delta(0)}{4.28~k_B}=$~87~K.

The simulated spectra can then be fitted with the same fitting method described above to identify the relevant temperature features and their parameter-based origin. Figure~S\ref{fig:figureS16}(e)(f)(g)(h) should be compared with Fig.~S\ref{fig:figureS10}(c)(d)(e)(b) respectively. The excellent agreement between the two fitting results - including the same fitting instability originated from the broad near-zero-energy peaks in the antinodal $\mathbf{k}_F$ EDCs near $T_\text{gap}$ - quantitatively confirms the gap opening temperature to be $\sim$ 20~K higher than the bulk $T_c$.

The spectral evolution of this fluctuating superconducting gap is also substantially different from that in the pseudogap region in optimally and underdoped cuprates. Figure~S\ref{fig:figureS10}(f)(g)(h)(i) show the same quantities discussed in Fig.~\ref{fig:figureS130}(c), Fig.~S\ref{fig:figureS10}(d)(e)(b), but derived near the antinode in an optimally doped Bi-2212 ($T_c$ = 98~K). Comparing to the fluctuating superconducting gap demonstrated here in overdoped Bi-2212, the ``pseudogap'' region in optimally doped sample is characterized by spectral gap larger than the superconducting gap, an order of magnitude larger spectral width (to an extent to destroy the quasiparticle all together), decreasing spectral peak intensity and zero-energy spectral weight with cooling.

\section{Temperature dependent high stats spectra in $p$ = 0.22 sample}

\begin{figure*}
	\captionsetup{width=2\columnwidth,justification=RaggedRight}
	\includegraphics[width=2\columnwidth]{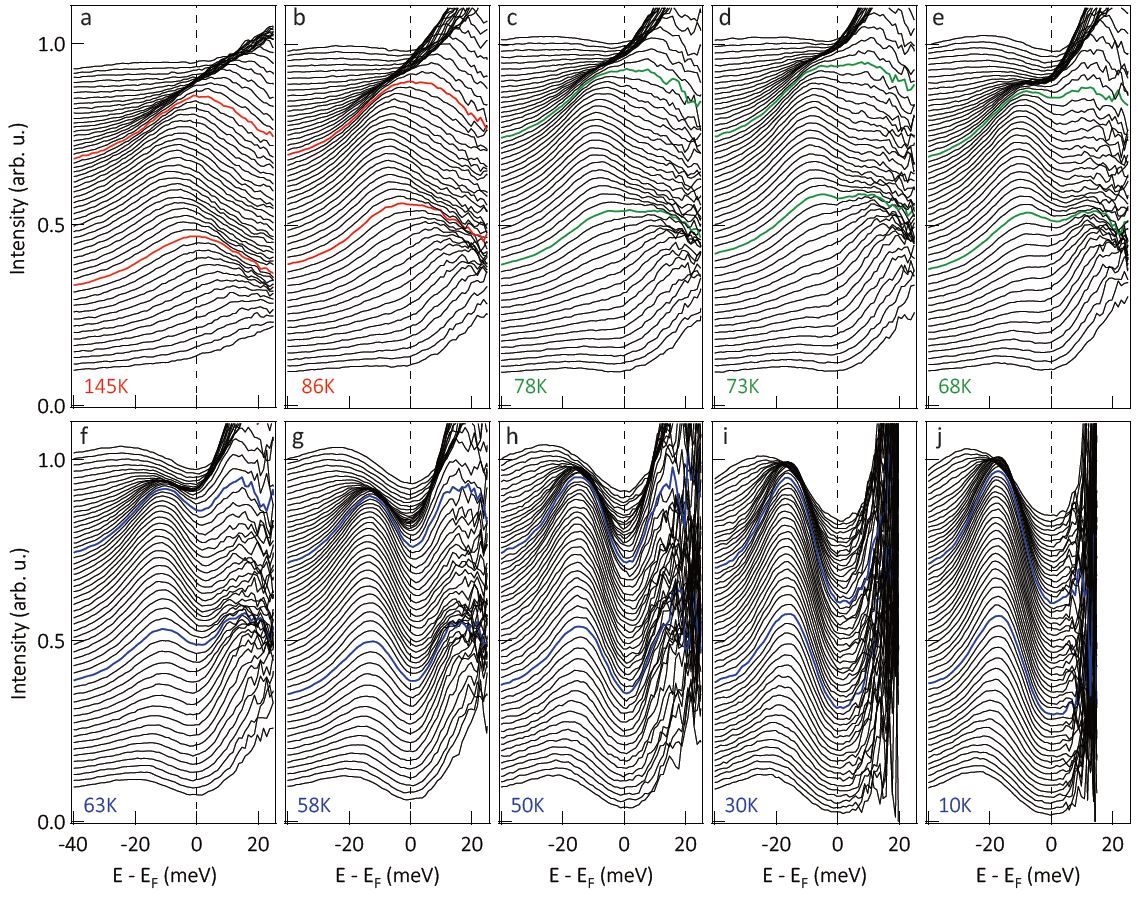}
	\caption{Raw temperature dependent Fermi function divided EDCs along the antinodal BZ boundary in overdoped Bi-2212 (p = 0.22, $T_c$ = 65~K). Red - normal state. Green - fluctuating temperature region. Blue - superconducting state. Colored lines highlights the EDCs at Fermi momenta.}
	\label{fig:figureS3}
\end{figure*}

Figure S\ref{fig:figureS3} shows the raw stacked antinodal EDCs after Fermi function division from 10~K to 145~K in overdoped Bi-2212 ($p$ = 0.22, $T_c$ = 65~K). The equal particle-hole mixing in the fluctuating region is clearly demonstrated from (b) to (e), contrasting the featureless broad spectra (on the order of hundreds of ~meV) that are usually considered one of the defining characters of the pseudogap~\cite{renner1998pseudogap,hashimoto2014energy}.

\section{Particle-hole symmetry in overdoped Bi-2212}

\begin{figure*}
	\captionsetup{width=2\columnwidth,justification=RaggedRight}
	\includegraphics[width=1.5\columnwidth]{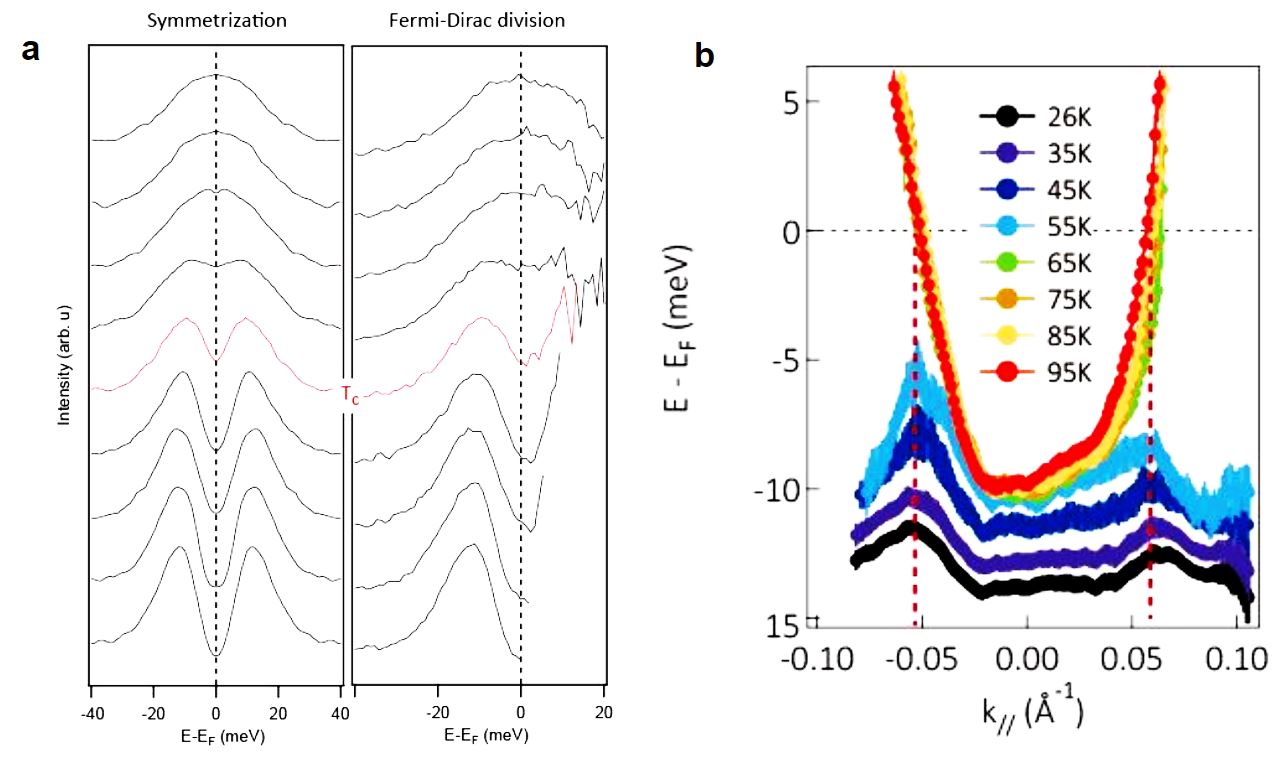}
	\caption{Particle-hole symmetry of the antinodal spectral gap in overdoped Bi-2212. (a) Temperature dependent EDCs at the antinodal $\mathbf{k}_F$ in p = 0.23 Bi-2212. Left: symmetrized EDCs. Right: Fermi function divided EDCs. Temperature step between each vertically offset curve is 10~K. (b) Fitted antinodal dispersion along the BZ boundary in $p$ = 0.24 Bi-2212 for various temperatures. The slight distortion in the center is due to a weak superstructure band contamination. Red dash denotes the alignment between the superconducting state quasiparticle dispersion back-bending momenta $\mathbf{k}_G$ and the normal state Fermi momenta $\mathbf{k}_F$.}
	\label{fig:figureS11}
\end{figure*}

\begin{figure*}
	\captionsetup{width=2\columnwidth,justification=RaggedRight}
	\includegraphics[width=1.7\columnwidth]{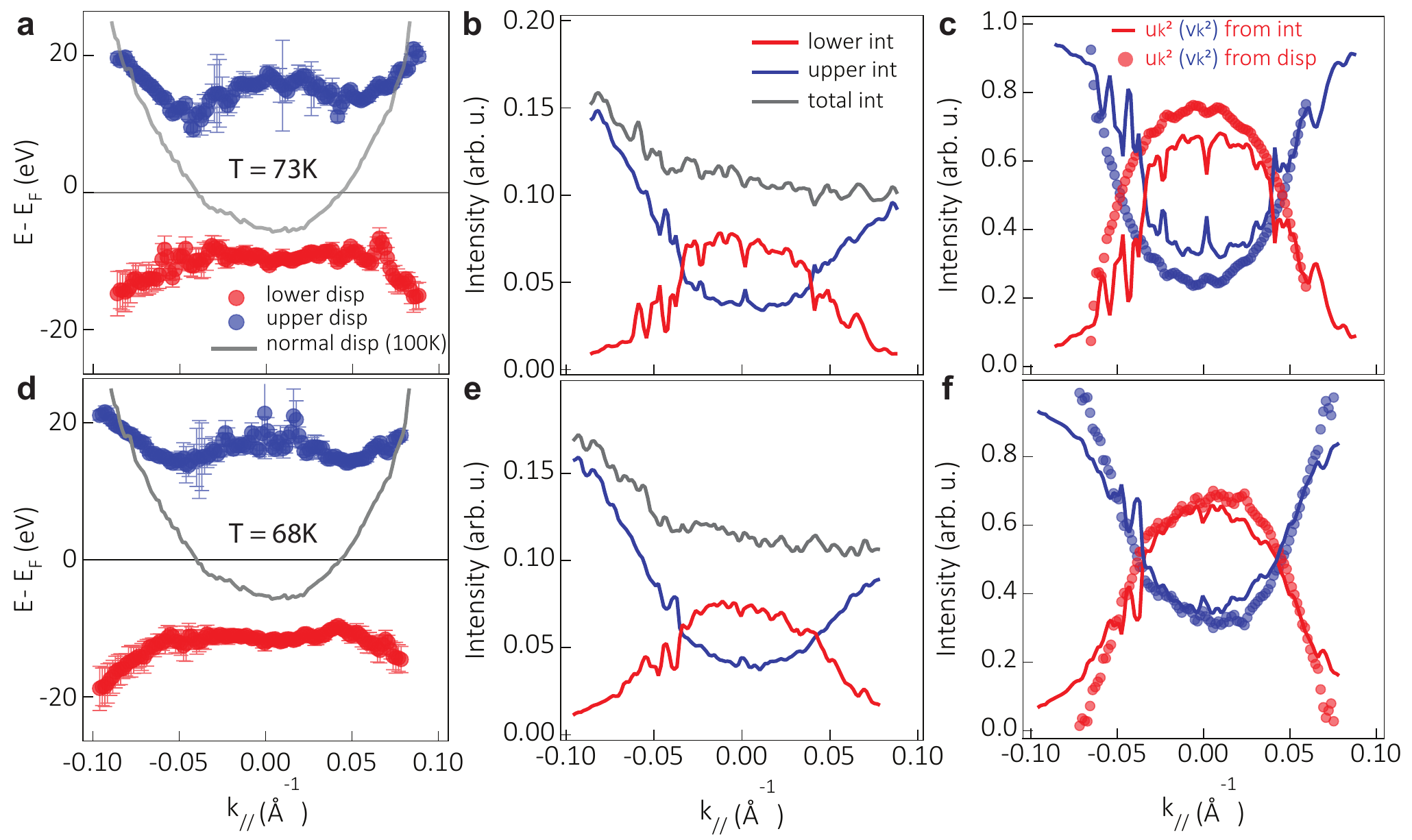}
	\caption{Extraction of $u_\mathbf{k}^2$ and $v_\mathbf{k}^2$ near the antinode at (a)-(c) 73~K and (d)-(f) 68~K. (a)(d) Gaussian fitted dispersions for both the upper (blue) and lower (red) branches of Bogoliubov quasiparticles. Grey curve is the normal state dispersion fitted at 100~K. (b)(e) Corresponding quasiparticle spectral intensities for the upper (blue) and lower (red) branches. Grey line is the sum of the red and blue curves, representing the matrix element effect induced momentum dependence in the intensity distribution. (c)(f) $u_\mathbf{k}^2$ (red) and $v_\mathbf{k}^2$ (blue) derived from the fitted intensity (line) and dispersion (circle) of the lower and upper branches. The grey curves in (b)(e) are used as the normalization factor to obtain the blue and red lines in (c) and (f) in order to satisfy $u_\mathbf{k}^2+v_\mathbf{k}^2=1$.}
	\label{fig:figureS18}
\end{figure*}

Other than directly observing the Bogoliubov quasiparticle dispersion of both the lower and upper branches, other indirect measures were also used in the literature to address the particle-hole symmetry of an energy gap. The first is through the examination of the Fermi function divided energy spectra at the single momentum - normal state $k_F$. It can be seen from Fig.~S\ref{fig:figureS11}(a) that the Fermi function divided EDCs agree with the symmetrized EDCs rather well, with the spectral gap minimum within 2~meV of the Fermi energy. This also emphasizes the importance of using Fermi function divided spectra, rather than symmetrized EDCs, to extract temperature dependence of energy gaps~\cite{hashimoto2010particle}. This is because when an energy gap no longer centers at the Fermi energy, symmetrization will incorrectly enforce particle-hole symmetry, and lead to artificially premature gap closure (gap center above $E_F$) or persistent gap (gap center below $E_F$).

When the unoccupied side of the spectra are difficult to obtain due to either lack of statistics or extremely low thermal population, one may also define a quantitative measure of particle-hole symmetry breaking along the momentum axis by the momentum shift $\Delta|\mathbf{k}_F| = |\mathbf{k}_F-\mathbf{k}_G|$, where $\mathbf{k}_F$ is the Fermi momenta and $\mathbf{k}_G$ is the gap back-bending momentum defined by the quasiparticle dispersion on the occupied side. Figure~S\ref{fig:figureS11}(b) plots the antinodal BZ boundary dispersion of an overdoped Bi-2212 (p = 0.24, $T_c$ = 49~K) at various temperatures. It is clearly seen that $\mathbf{k}_F$ and $\mathbf{k}_G$ align over the temperature range probed. Comparing to the pseudogapped state in Bi-2201~\cite{hashimoto2010particle}, the particle-hole symmetry is restored in sufficiently overdoped Bi-2212.

\section{Superconducting spectral function}

To quantify the particle-hole mixing near the antinodal region, we use the following relations for the superconducting state spectral function:
\begin{equation}
\begin{split}
    A_{\mathbf{k},\omega} &= u^2_\mathbf{k} A^\text{particle}_{\mathbf{k},\omega} + v^2_\mathbf{k} A^\text{hole}_{\mathbf{k},\omega}\\
\end{split}
\end{equation}
\noindent where $u_\mathbf{k}^2 (v_{\mathbf{k}}^2) $ are the probability amplitudes for the Bogoliubov quasiparticle to be a particle (hole), $A^\text{particle}_{\mathbf{k},\omega}$ and $A^\text{hole}_{\mathbf{k},\omega}$ are the particle and hole parts of the superconducting spectral function. At any given momentum, the intensity ratio between the lower and upper Bogoliubov branches is $u^2_\mathbf{k}/v^2_\mathbf{k}$. This provides an independent experimental measure of $u_\mathbf{k}^2/v_\mathbf{k}^2$ besides those inferred from the quasiparticle dispersion (Fig.~S\ref{fig:figureS18}) via:
\begin{equation}
    u^2_\mathbf{k}, v^2_\mathbf{k} = \frac{1}{2}\big(1 \pm \frac{\epsilon_\mathbf{k}}{\sqrt{\epsilon_\mathbf{k}^2+\Delta_\mathbf{k}^2}}\big)
\end{equation}

The measured spectral intensity will still be modulated by a superimposing dipole-transition matrix element $|M(\mathbf{k})|^2$, which will be normalized out in Fig.~S\ref{fig:figureS18}.

\end{document}